\documentclass[journal]{IEEEtran}
\usepackage{caption}
\usepackage{algorithmic}
\usepackage[linesnumbered,ruled]{algorithm2e}
\usepackage{graphics,epstopdf}
\usepackage{balance}
\usepackage{listings}
\usepackage{comment}
\usepackage{diagbox}
\usepackage{enumitem}
\usepackage{xcolor}
\usepackage{subfig}
\usepackage{xcolor}
\usepackage{fixmath}
\usepackage{cite}
\usepackage{amsmath,amssymb,amsfonts}
\usepackage{graphicx}
\usepackage{textcomp}
\usepackage{tabularx}
\usepackage{booktabs}
\usepackage{longtable}
\usepackage{multirow}
\usepackage{colortbl}
\usepackage{hhline}
\usepackage{diagbox}
\usepackage[
  separate-uncertainty = true,
  multi-part-units = repeat
]{siunitx}
\usepackage{tabularx,colortbl}
\usepackage{stmaryrd}
\usepackage{hyperref} % for autoref
\usepackage{cite}
\usepackage{enumitem}
\usepackage{soul}
\usepackage{bm} %bold math \bm{s}
\usepackage{mathtools} % text on arrow
\usepackage{mathrsfs} % for A as adversary
\usepackage{multirow}
\usepackage{array}
\ifCLASSINFOpdf

\else

\fi

\begin{document}

\title{Privacy-Preserving and Efficient Data Collection Scheme for AMI Networks Using Deep Learning}
%Countering Presence Attacks in Efficient Advanced Metering Infrastructure using Deep Learning

\author{\IEEEauthorblockN{Mohamed I. Ibrahem,
Mohamed Mahmoud,~\IEEEmembership{Senior~Member,~IEEE,}
Mostafa M. Fouda,~\IEEEmembership{Senior~Member,~IEEE,}\\
Fawaz Alsolami,~\IEEEmembership{Member,~IEEE,}
Waleed Alasmary,~\IEEEmembership{Senior~Member,~IEEE,}\\
and Xuemin (Sherman) Shen,~\IEEEmembership{Fellow,~IEEE.}}
\vspace{-0.3in}

\thanks{Corresponding author: Mohamed I. Ibrahem.}
\thanks{M. I. Ibrahem, and M. Mahmoud are with the Department of Electrical and Computer Engineering, Tennessee Tech. University, Cookeville, TN 38505 USA (e-mail: miibrahem42@tntech.edu; mmahmoud@tntech.edu).}
\thanks{M. M. Fouda is with the Department of Electrical and Computer Engineering, College of Science and Engineering, Idaho State University, Pocatello, ID 83209, USA (e-mail: mfouda@ieee.org).}
\thanks{F. Alsolami is with the Department of Computer Science, King Abdulaziz University, Jeddah, Saudi Arabia (e-mail: falsolami1@kau.edu.sa).}
\thanks{W. Alasmary is with the Department of Computer Engineering, Umm Al-Qura University, Saudi Arabia (e-mail: wsasmary@uqu.edu.sa).}
\thanks{X. Shen is with Department of Electrical and Computer Engineering, University of Waterloo, Waterloo, Ontario, Canada (e-mail: sshen@uwaterloo.ca).}
%\vspace{-1cm}
}

\maketitle

% The paper headers
\markboth{Ibrahem \MakeLowercase{\textit{et al.}}: Privacy-preserving and Efficient Data Collection Scheme for AMI networks Using Deep Learning}%
{}
\IEEEpeerreviewmaketitle
\begin{abstract}

In advanced metering infrastructure (AMI), smart meters (SMs), which are installed at the consumer side, send fine-grained power consumption readings periodically to the electricity utility for load monitoring and energy management. Change and transmit (CAT) is an efficient approach to collect these readings, where the readings are not transmitted when there is no enough change in consumption.
However, this approach causes a privacy problem that is by analyzing the transmission pattern of an SM, sensitive information on the house dwellers can be inferred. For instance, since the transmission pattern is distinguishable when dwellers are on travel, attackers may analyze the pattern to launch a presence-privacy attack (PPA) to infer whether the dwellers are absent from home. 
In this paper, we propose a scheme, called ``STDL'', for efficient collection of power consumption readings in AMI networks while preserving the consumers' privacy by sending spoofing transmissions (redundant real readings) using a deep-learning approach. We first use a clustering technique and real power consumption readings to create a dataset for transmission patterns using the CAT approach. Then, we train an attacker model using deep-learning, and our evaluations indicate that the success rate of the attacker is about 91\%. Finally, we train a deep-learning-based defense model to send spoofing transmissions efficiently to thwart the PPA.
Extensive evaluations are conducted, and the results indicate that our scheme can reduce the attacker's success rate, to 13.52\% in case he knows the defense model and to 3.15\% in case he does not know the model, while still achieving high efficiency in terms of the number of readings that should be transmitted.
Our measurements indicate that the proposed scheme can reduce the number of readings that should be transmitted by about 41\% compared to continuously transmitting readings.

%%%%%%%%%%%%%%%%%%%%%%%%%%%%%%%%%%%%%%%%%%%%%%%%%%%%%%%%%%%%%%%%%%%%%%

\end{abstract}

\begin{IEEEkeywords}
Privacy preservation, smart grid, AMI networks, traffic analysis attack
\end{IEEEkeywords}

\vspace{-0.1in}

\section{Introduction}

Smart grid is a revolutionary upgrade to the traditional power grid that aims to increase the reliability of electricity delivery, optimize grid operation, and reduce the emissions of greenhouse gases by using more renewable resources for power generation~\cite{6298960}. 
Advanced metering infrastructure (AMI) is one of the most important components of the smart grid.
AMI enables bi-directional communication between the smart meters (SMs), which are deployed at consumer premises, and the electricity utility (EU) to collect fine-grained power consumption readings periodically (every few minutes) to obtain detailed information necessary for monitoring the grid load and managing the energy supply efficiently~\cite{5741147,5622050}. 

However, revealing the consumers' fine-grained power consumption readings creates a serious privacy problem. This is because the readings can be analyzed to infer sensitive information about the consumers' life habits such as the appliances that are being used, when consumers sleep, when they return home, etc~\cite{issue1}. This is because each appliance has a unique signature on the power consumption.
To preserve consumers' privacy, encryption schemes have been used to hide the fine-grained readings while enabling the EU to use the encrypted readings for different purposes such as load monitoring~\cite{7841782,ISNCC20,ibrahem2020efficient}.

%add my PMBFE conf paper here
%li2011secure

On the other hand, reporting the fine-grained readings periodically results in transmitting a massive amount of data by each SM. What makes the problem worse is that the AMI networks are scalable, i.e., they have millions of SMs. This leads to inefficient use of the available communication bandwidth, and in case of using cellular networks to transmit the SMs’ readings, it is costly to transmit this large amount of data, particularly after considering the overhead caused by the encryption schemes. 
In a more efficient power consumption readings collection approach, called change and transmit (CAT), SMs do not need to report their consumption when there is no enough change compared to the last reported reading~\cite{Dr_Li}. 
More specifically, in CAT approach, a reading is transmitted when the absolute value of the percentage of change in the consumption exceeds a predefined threshold
% However, using CAT approach results in a trade-off between the efficiency in terms of the number of transmitted readings and error in terms of the difference between the actual power consumption and the reported reading.
% , i.e., a smaller threshold leads to transmitting more number of readings but with less error.
% More specifically, in CAT approach, a reading is transmitted when the absolute value of the percentage of change in the consumption exceeds a predefined threshold, so the bigger the threshold, the greater the error, i.e., the difference between the real consumption and the last reported reading.
% Luckily, as will be investigated in this paper, the EU needs the total consumption of an AMI network for load monitoring and energy management~\cite{7841782,ISNCC20,ibrahem2020efficient}, and because the error in the readings can be positive (when a reading is more than the last reported one) or negative (when a reading is less than the last reported one), the error may cancel each other after aggregating the readings, and hence, the aggregated reading error is reduced.
% The selection of the threshold is based on the sensitivity of the application to the accuracy of the readings~\cite{WRIGHT2007389}.
% Hence, in an efficient power consumption readings collection approach, called change and transmit (CAT), an SM transmits the power consumption reading only when the absolute difference between the current consumption and the last reported consumption reading exceeds a predefined threshold.

Nevertheless, using the CAT approach causes a privacy problem that is by analyzing the transmission patterns of the SMs, sensitive information on the consumers can be revealed even if the readings are encrypted~\cite{5403146,Dr_Li}. For instance, these information can include whether the house dwellers are on travel (absent) or present at home, sleeping cycles, meal times, the number of dwellers, etc~\cite{7093120,weaver2014perspective}, because these events have distinguishable impact on the transmission pattern.
% Also, if the attacker observes a lack of power consumption change events during a long period, then, it can be inferred that the dwellers are not at home with high confidence.
% Furthermore, the attacker can also gain information about the consumers' activities such as whether they are awake or asleep, and cooking or having meals, and other activities. 
These private information can be used for criminal purposes such as finding the right time to commit crimes, e.g., kidnapping, pillage, burglary, arson, and vandalism~\cite{7066595}.
% , and thus brings significant jeopardy to the power user. On the other side, the consumers' transmission pattern may be sold to or monitored by insurance companies or competitor suppliers, whose primary goal is to make profits, to adapt their plans based on the consumers' transmission pattern. Such information should be hidden to ensure fair trade market~\cite{8382299}. 

In this paper, we focus on thwarting presence privacy attack (PPA) that aims at inferring whether the house dwellers are present at home or absent by analyzing the transmission pattern of the SM.
% ; however, our methodology can be extended to detect other attacks, as we gave some examples, that are caused by analyzing the consumers' transmission patterns. 
% For instance, when the power consumption activity is more intensive, the power consumption change will also be increased, which results in more packets due to the policy of transmission. Hence, the eavesdropper can easily determine the current power consumption intensity. Then, for a typical period of intensive power consumption such as dinner cooking time, if the eavesdropper finds that the number of packets is significantly lower than the normal level, it is highly possible that the house owner is not at home.
% Such information should be hidden from competitors to ensure fair electricity trade market [33], [34].
% In PPA, an attacker analyzes the transmissions of an SM to learn whether the house dwellers are on travel (i.e., absent) or not (i.e., present). 
The seriousness of PPA is due to the following facts. First, PPA can be launched in an undetectable way since attackers usually work completely in the receiving mode to overhear the transmissions without disrupting the communications. Second, the use of encryption schemes cannot thwart the attack because the attackers analyze the transmission patterns and do not need to decrypt the readings. Third, AMIs mostly use wireless communication~\cite{6177682,Dr_Li}, and the broadcast nature of this communication can facilitate collecting transmission patterns of victim consumers. In summary, the research problem we address in this paper is \textit{how to thwart PPA, while achieving a high efficiency in terms of the number of readings that should be transmitted}.
% , i.e., no one should learn the absence of the house dwellers by observing their transmission pattern} to preserve privacy.

In the literature, a countermeasure to PPA has been proposed in~\cite{Dr_Li} by sending spoofing transmissions (packets for redundant readings) to make it difficult for the attacker to analyze the transmission pattern to infer whether dwellers are absent or present. 
Artificial spoofing packets (ASP) are generated either by using the distribution of the power consumption transmission pattern of the consumers or using their old transmission patterns.
% , where the attacker monitors the consumers' transmission pattern to infer his/her presence at home. 
% Two heuristic-based defense methods have been proposed %, namely the Poisson packet generation and history template (HT) based generation, 
% to generate artificial spoofing packets to deceive the attacker. 
% according to a heuristic-based defense scheme.% Li called his model as mathematical model in his conclusion.
% In Poisson packet distribution, the memory can store the average number of power consumption changes within a time period in different time intervals when the consumer is present at home. Therefore, the number of changes can be used to generate spoofing packets when the consumer is not at home so that the total number of changes within a certain time interval equal to that number of changes stored in the memory. 
% On the other hand, the history templates, which are stored in the memory, can also be used to generate the spoofing packets by either repeating the same previous transmission patterns or randomly generate the spoofing packets according to the empirical distribution of the time intervals between significant power consumption changes.
However, the scheme suffers from the following limitations. First, the success rate of the attackers is too high, i.e., above 60\%. 
% , which means the attacker can infer, with high confidence, whether the house dwellers are at home or not.
Second, the attacker model is very simple, i.e., the attacker collects transmission patterns and compares their distribution with the distribution of the old transmission patterns of the consumers. Third, the scheme in~\cite{Dr_Li} considers limited data that is for one household over only 33 days. Fourth, the utility is assumed to be trusted to learn the readings and identify the spoofing transmissions. 
% and the defense is limited only to the eavesdroppers. 
Finally, the proposed defense in~\cite{Dr_Li} is customized for one consumer because it depends on the distribution of the transmission pattern of the consumer.

% The K-S test compares between two distributions and returns the error. Hence, if the error does not exceed a predefined threshold, the two distribution are close to each. Once the attacker finds that packets actually contain many spoofing ones, it can determine that the house owner is on leave.
% Li et al.~\cite{Dr_Li} suggested that the SM send the power consumption reading if and only if the reading exceeds a predefined threshold, i.e., (CAT), to improve the power consumption of the resources or the spectrum efficiency. However, When a change occurs, the eavesdropper can launch PPA by monitoring the radio activity of a certain power consumer, and hence the privacy of the consumer is violated, i.e., the eavesdropper can infer whether the consumer is at home. 

Therefore, in this paper, we address these limitations by proposing a new scheme, called ``STDL'', for efficient collection of power consumption readings while thwarting PPA by transmitting \textbf{s}poofing \textbf{t}ransmissions (redundant real readings) using \textbf{d}eep \textbf{l}earning. 
First, a clustering technique and real power consumption readings are used to create a dataset for transmission patterns using CAT approach. Then, an attacker model is trained using deep-learning, and our evaluations indicate that the success rate of the attacker is about 90\%. Hence, to thwart PPA, we train a deep-learning-based defense model to send spoofing transmissions (redundant real readings) efficiently. This defense model is trained on the transmission pattern of the consumers when they are present at home. Next, the defense model is used to generate spoofing transmissions so that the attacker cannot distinguish the pattern generated by the model from the patterns transmitted when the consumers are at home to preserve consumers' privacy. 

% Since there is no ready-to-use dataset that contains the information about the consumers' presence status, we developed two approaches to label the dataset to use the generated dataset for training and evaluating STDL. Using a real power consumption dataset, we evaluated the performance of STDL, and the results indicate that our scheme can thwart PPA while reducing the communication bandwidth by about 41\% compared to sending consumption periodically.
The evaluations of our scheme indicate that our scheme can significantly reduce the attacker's success rate to less than 10\% compared to more than 60\% using the proposed scheme in~\cite{Dr_Li}. Our scheme is also robust even if the attacker knows the defense model. In addition, it can reduce the number of transmitted readings by about 41\% compared to sending consumption readings periodically.
% Moreover, unlike~\cite{Dr_Li}, our scheme does not need high memory storage to store a large number of historical records to be used in generating spoofing packets.

Our contributions in this paper can be summarized as follows.
\begin{itemize}
    \item To the best of our knowledge, this work is the first to investigate PPA using a deep-learning approach. Compared to the literature, we consider a more sophisticated and strong attack model that is trained on the consumers' transmission patterns to detect whether the dwellers are at home or not. 
    
    \item We propose a defense model against PPA which offers a fairly low attacker's success rate compared to the literature since it is based on a deep-learning approach which results in generating patterns that look like the patterns transmitted when the consumer is at home.
    
    \item Our defense model is general (one model for all consumers) and it can be applied to new consumers who have no old transmission patterns.
    
    \item We assume that the network nodes, including the SMs, EU, and aggregator, are honest but curious, while the literature assumes that they are trusted and the defense is limited only to the eavesdroppers.
    
\end{itemize}

% On the other hand, the design of our defense scheme is based on a deep-learning model which is trained on the transmission patterns when the consumers are present at home to generate spoofing transmissions so that the attacker cannot distinguish the pattern generated using the defense model from the patterns transmitted when the consumers are at home to preserve consumers' privacy. 
% the design of our defense scheme is based on a deep learning approach, which usually performs better than heuristic-based approaches~\cite{jokar2016electricity,9148937}.

% Besides, we assume that the internal nodes, e.g., the utility and aggregator, are honest but curious.

% Although STDL is proposed to mitigate PPA, but it can be extended to other problems. 
% Finally, in this paper, we explain how EU can monitor the grid load while thwarting PPA and improve bandwidth efficiency.
% by adding the encrypted power consumption readings from all the consumers

% in Li's paper, the authors assumed that the attacker cannot decode the packets and did not explain how to do that and 
% In the CAT policy used by Li, the measurement is transmitted only if at some point the difference between two consecutive measurements exceeds the threshold and may, therefore, suffer from large tracking errors. But, in our case, we compared the current value with the last reported value, and hence the error is negligble as we will see in Fig~\ref{}.
% , and proved that even if the attacker knows the defense, she cannot distinguish between spoofed and real patterns.

The remainder of this paper is organized as follows. Section~\ref{Related_Work} discusses the related works. Then, our system models and design objectives are discussed in Section~\ref{sec:System Models}. Section~\ref{sec:Preliminaries} illustrates the preliminaries used in this paper. The datasets created for training our models are presented in Section~\ref{dataset_preprocessed}.
Our envisioned defense scheme is presented in Section~\ref{PS}. Next, the performance evaluation of our scheme is discussed in Section~\ref{PE}. Finally, the paper is concluded in Section~\ref{conclusion}.

\section{Related Work}\label{Related_Work}
%%%%%%%%%%%%%%%%%%%%%%%%%%%%%%%%%%%%%%%%%%%%%%%%%%%%%%%%
Most of the existing privacy-preserving electricity consumption readings collection schemes in the literature~\cite{7841782,ISNCC20,ibrahem2020efficient} assume that the smart meters report the power consumption readings periodically even if there is no enough change in the consumption compared to the last reported reading.
These schemes use privacy-preserving data aggregation schemes to aggregate the encrypted readings and allow the EU to obtain the aggregated reading for load monitoring and energy management without being able to access the individual readings of the SMs to preserve the consumers' privacy.
% While some studies are based on Paillier cryptosystem to allow performing mathematical operations over encrypted data without decrypting it~\cite{7959204,7348648,7497514,8633950}, other studies use different approaches as follows.

Mohammed \textit{et. al.}~\cite{7841782} proposed a privacy-preserving data collection scheme that uses one-time secrets to mask the readings so that the aggregated reading is obtained by adding the masked readings of the AMI network. This scheme cannot be applied in case of CAT approach since it requires each SM to use a new mask in every data collection cycle, so if an SM does not send a reading, the last reported masked reading cannot be used in the aggregation. 
Moreover, a functional encryption-based scheme is proposed in~\cite{ISNCC20,ibrahem2020efficient} for aggregating the consumers' readings while preserving the consumers' privacy.
This scheme cannot also be used in case of CAT approach since it cannot use the last power consumption reading of an SM when the consumption does not change because the scheme requires using a fresh timestamp in each reading.
% if an SM does not send a power consumption reading, its last reported reading should be added to the aggregated reading, and it is difficult to do that
% the last reported reading by an SM
% , but it cannot add the previously stored readings because the aggregation process depends on a timestamp. 
% Fan \textit{et. al.}~\cite{6578183} proposed a privacy-preserving data aggregation scheme that uses blinding factors to hide the power consumption readings from internal attacks only, e.g., electricity suppliers. However, to ensure the privacy of the consumers, the power consumption readings should be hidden from both external and internal adversaries.
Hence, the schemes proposed in~\cite{7841782,ISNCC20,ibrahem2020efficient} focused on preserving consumers' privacy by hiding the power consumption readings, while our focus in this paper is on preserving the consumers' privacy by hiding the information that can be inferred from the consumers' transmission patterns in case of using CAT approach.

% this scheme did not take external attackers into account. a privacy-enhanced data aggregation scheme is proposed against internal attackers in smart grid. In this scheme, electricity suppliers can learn about the current energy usage of each neighborhood to arrange energy supply and distribution without knowing the individual electricity consumption of each user.  privacy-preserving approach designed to mitigate risks due to power data aggregation from internal attacks in a smart grid. To ensure the privacy of the electricity consumption data, a P2DA scheme should provide privacy, i.e., neither external adversaries nor internal adversaries should be able to access a specific consumer’s electricity consumption data. Fan et al.’s scheme suffers from the key leakage problem because any adversary could extract the user’s private key through the public information.
% and also learn the electricity consumption of consumers. The work does not take external attackers into account.
% , He et al~\cite{8633950} tried to add a random Gaussian noise to the SM's consumption reading so that it is infeasible for adversaries to get the original consumption data while allowing the EU to recover the total consumption reading of all the SMs.

%%%%%%%%%%%%%%%%%%%%%%%%%%%%%%%%%%%%%%%%%%%%%%%%%%%%%%%%%%%%%%%%%%%%%%%%%%%%%%%%%%
% Finally, another subsection for non-periodic transmission and mention the papers that worked on it and mention that there is no one concern privacy except Li's Paper and talk about its work.

Few works in the literature have investigated CAT approach in smart grid~\cite{6521385,CARRIEARMEL2013213,7362875,7229338,7007667,7875140,Dr_Li}.
% In this approach, an SM transmits a power consumption reading only if it differs by more than a predefined threshold from the previous reading.
Some of these works studied using CAT approach in different applications such as demand/response~\cite{6521385} and load disaggregation~\cite{CARRIEARMEL2013213} that tries to extract the power patterns of individual appliances from the aggregated power pattern measured by the SM to analyze the power consumption of each appliance. 
% Load disaggregation is beneficial to the consumers to determine which appliances are the most energy consuming ones, which appliances are faulty, and when it is time to retrofit old appliances; also, it is beneficial to the energy utilities to better forecast demand.
While other works~\cite{7229338,7362875,7007667,7875140} investigate finding a good value for the change in the consumption that necessitates transmitting a reading.
% threshold in which the difference between the reported and true readings is acceptable. 
% , e.g., load forecasting with erroneous data may lead to generating more or less electricity than is actually needed~\cite{WRIGHT2007389}. 
% This is because there is a trade-off between the transmission rate and error, i.e., a smaller threshold leads to a higher transmission rate and lower error.
The schemes in~\cite{6521385,CARRIEARMEL2013213,7362875,7229338,7007667,7875140} that use CAT approach do not consider consumers' privacy, so they are vulnerable to PPA. 

Li \textit{et. al.}~\cite{Dr_Li} have investigated PPA using CAT approach.
A defense scheme, called ``ASP'', is proposed to thwart PPA by generating artificial spoofing transmissions.
To generate the spoofing transmissions, two heuristic-based methods have been proposed, namely the Poisson packet generation and history template based generation.
% Simply, These transmissions can be generated using either the distribution of the power consumption changes of the consumers or their history templates that contain the historical transmission patterns.
In Poisson packet generation method, the SM's memory stores the average number of power consumption changes in different time intervals when the dwellers are present at home. Then, when the dwellers are not at home, spoofing transmissions are generated so that the total number of changes within a certain time interval equals to that stored in the memory. 
On the other hand, in the history template method, old transmission patterns are used to generate the spoofing transmissions by either repeating the old transmission patterns or randomly generating the spoofing transmissions according to the distribution of the time intervals between the transmissions.
% The generation of the spoofing transmissions is based on either the distribution of the power consumption changes or the consumers' historical records of the power consumption changes.
However, ASP suffers from the following limitations.

\begin{enumerate}
    \item The attacker's success rate is significantly high. In one experiment, it is assumed that the attacker knows a set of old transmission patterns, and the attacker's success rate in this case is above 60\%. The attacker uses a K-S test to compare the distribution of the transmission pattern that is received from an SM with the distribution of the transmission patterns that he/she knows. The K-S test is used to calculate the error between two distributions to determine how they are close to each other. 
    Another experiment assumes that the attacker does not have old transmission patterns, and in this case, the attacker collects transmission patterns sent in different time intervals and compares their distributions using the K-S test. If the attacker finds that the error between the distributions is small enough, he/she can conclude that the transmissions are spoofing, and thus the consumer is on travel. The experimental results demonstrate that the attacker's success rate is above 40\%. 
    Overall, these results indicate that the attacker can detect, with high confidence, whether the consumer is present or on travel
    % the pattern generated using ASP from the real one due to CAT approach.  Once the attacker finds that transmissions actually contain many spoofing ones, he/she can determine that the house dwellers are not at home. In addition, ASP requires a huge memory to store the old of power consumption changes of the consumers.% to obtain a better defense performance.
    
    \item The proposed attack mechanism is very simple, i.e., the attacker collects transmission patterns received from SMs and analyzes them by comparing their distribution with the distribution of the old transmission patterns by using the K-S Test. 
    Therefore, a more sophisticated attack mechanism may result in higher success rate.
    % Moreover, ASP is based on heuristic-based approaches which do not perform well in such problems~\cite{jokar2016electricity,9148937}.

    \item The study is very limited since only one household is considered over only 33 days. Also, the proposed scheme is customized, i.e., can be used for only one consumer, because it generates the spoofing transmissions based on the distribution of the transmission patterns of each consumer. 
    % follow/learn
    
    \item It is assumed that the utility is trusted. In practice, it is impossible to guarantee that a party (which is assumed trusted) does not misuse the captured data of the consumers.
    
\end{enumerate}

%%%%%%%%%%%%%%%%%%%%%%%%%%%%%%%%%%%%%%%%%%%%%%%%%%%%%%%%%%%%%%%%%%%%%%%%%%%%%%%%%%
%Second, another subsection for traffic analysis and this can be taken from Hammad's paper. They are different stories.

In wireless sensor networks (WSNs), spoofing transmissions have been used to counter traffic analysis attacks that aim to determine the location of the sender node and the sink~\cite{7479541,7289106}. However, these schemes cannot be applied to counter transmission pattern analysis in the AMI networks because WSNs have different characteristics and the privacy problem is different. 
We can summarize the main differences as follows: ($1$) The proposed schemes aim to preserve the location privacy of the source/destination sensor nodes, but we aim to preserve the privacy of the consumers' activities in AMI networks; and ($2$) In WSNs, there is a restriction on energy consumption because most of the nodes are battery powered, but this restriction does not exist in AMI networks.

\section{System Models and Design Objectives}  \label{sec:System Models}

This section discusses the considered network and threat models. 
%%%%%%%%%%%%%%%%%%%%%%%%%%%%%%%%%%%%%%%%%%%%%%%%%%%%%%%%

\subsection{Network Model} \label{Network_Model_sec} 
\begin{figure}[t]
\centering
\includegraphics[width=3.2in]{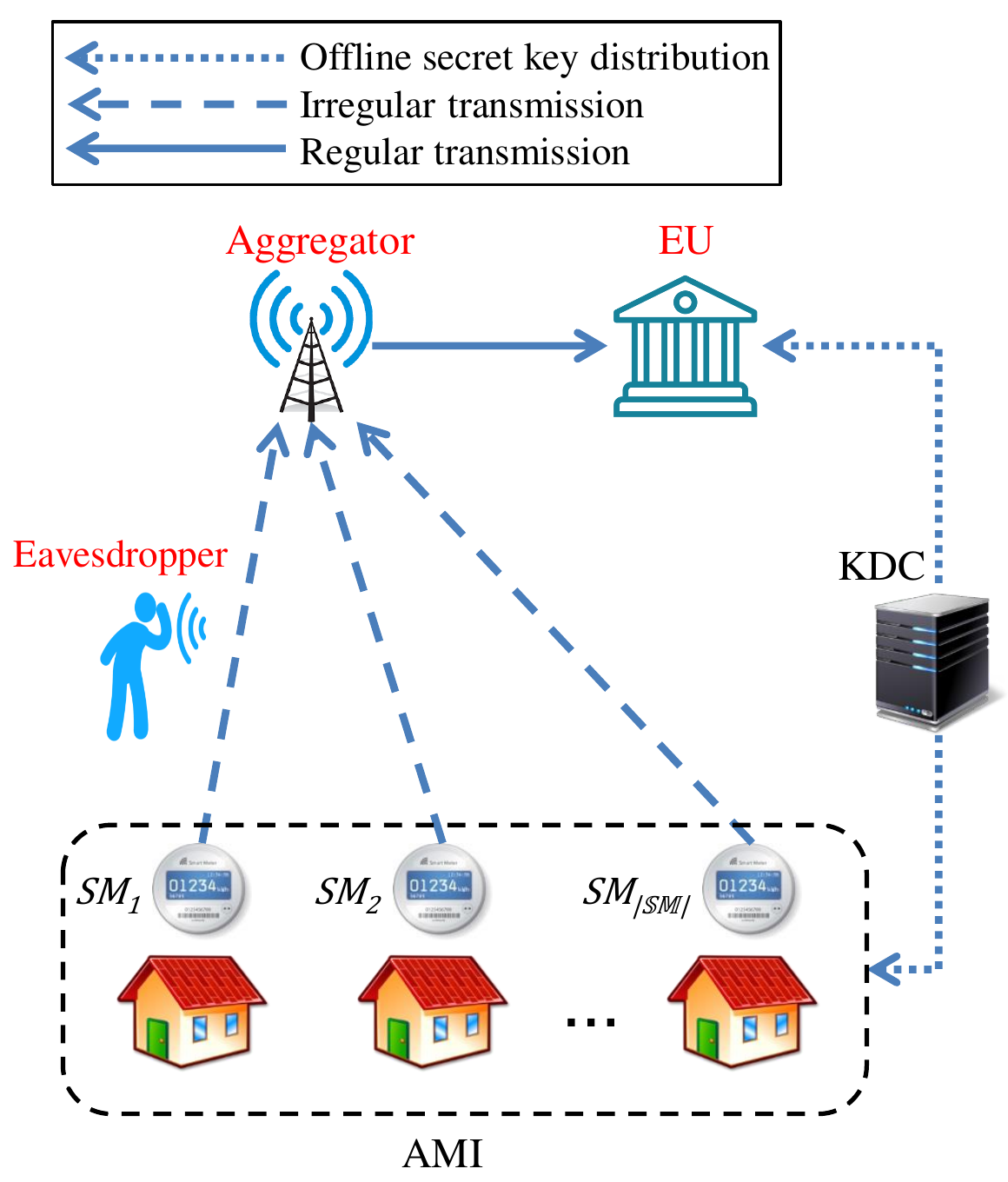}
\caption{Network Model.} \label{System_model}
\end{figure}

As shown in Fig.~\ref{System_model}, our considered network model includes SMs, EU, aggregator, and an offline key distribution center (KDC). The role of each entity is described as follows.
\begin{itemize}

    \item \textit{Key Distribution Center (KDC)}: It is responsible for generating and distributing the public and private keys of the cryptosystem used in our scheme. After key distribution, the KDC is not involved in the SMs' readings collection process. 
    % KDC can be operated by an authority such as the Department of Energy (DoE). 
    
    \item \textit{Smart Meter (SM)}:
    A set of SMs, $\mathbb{SM}=\{SM_i, 1 \leq i \leq |\mathbb{SM}|\}$, form an AMI network, where each $SM_i$ reports the real-time fine-grained power consumption reading if and only if there is an enough change in the power consumption comparing to the last reported reading, i.e., if the absolute value of the percentage of the change between the current consumption and the last reported reading exceeds a certain threshold. For instance, if the threshold is 5\%, a reading is sent when the change in the consumption is above +5\% or below -5\%.
    Also, each $SM_i$ encrypts its power consumption reading and sends the ciphertext to a gateway, called the aggregator.
    % When a consumer leaves, he/she triggers the RSPDL defense scheme to determine whether it needs to send a redundant reading to the aggregator to prevent PPA. 
    % These readings are sent to the EU through a gateway, called a local aggregator. 
    In this case, the SMs may communicate directly with the gateway, or multi-hop data transmission is used to connect the SMs to the gateway, where some SMs may act as routers to relay other SMs' readings.

    \item \textit{Aggregator}: It acts as a local collector that aggregates all the received/stored encrypted power consumption readings sent by the SMs. Due to using the CAT approach, the aggregator stores the SMs' power consumption readings to reuse the last reported readings for future reporting cycles in case that an SM does not report a power consumption reading. After collecting the encrypted SMs' readings, the aggregator computes the ciphertext of the aggregated reading and forwards it to the EU without being able to reveal the individual readings or even the aggregated reading to preserve privacy.

    \item \textit{Electricity Utility (EU)}:
    The EU uses the fine-grained power consumption readings sent by SMs for load monitoring and energy management. It can use its decryption key to decrypt the ciphertext sent by the aggregator to obtain the total power consumption of the consumers of an AMI network while hiding the fine-gained power consumption reading of each consumer to preserve privacy.  
    
\end{itemize}

%%%%%%%%%%%%%%%%%%%%%%%%%%%%%%%%%%%%%%%%%%%%%%%%%%%%%%%%

\subsection{Threat Model}\label{subsec:Threat Model} 
Threats may arise from either internal or external attackers.
Internal attackers, which could be the SMs, the EU, or the aggregator, are assumed to be honest-but-curious. They follow the scheme honestly and do not disrupt the communication, but they are curious to infer whether the house dwellers are present at home or not by analyzing their transmission pattern.
% We also assume that an internal attacker can compromise an entity in our network model, e.g., EU or aggregator, and monitor the power consumption changes pattern to learn their habits and behaviors.
On the other hand, as shown in Fig.~\ref{System_model}, an external adversary $\mathcal{A}$, e.g., eavesdropper, can intercept the wireless transmissions between the SMs and the aggregator over a period of time to create the transmission patterns of the consumers to launch PPA. This is because AMIs mostly use wireless communications~\cite{6177682,Dr_Li}, and the broadcast nature of these communications can facilitate intercepting signals. 
Also, the attackers are assumed to have old absent/present records of the consumers' transmission patterns that can be used to train a deep-learning model to launch PPA.
% This model can be used to launch PPA where the collected transmission pattern is the input to the model, and the output is either the house dwellers are present at home or absent.

% The seriousness of PPA is that it can be launched in an undetectable way because attackers usually work completely in the receiving mode to overhear the transmissions without disrupting the communications. Furthermore, the use of encryption schemes cannot thwart PPA, i.e., the attacker does not need to decrypt the readings sent by the SMs.

Basically, the objective of this paper is to preserve the privacy of the consumers' activities by hiding the information that can be inferred from the consumers' transmission patterns, i.e., no one including the EU and aggregator shall be able to learn the absence of the house dwellers. This complements the works in the literature that focus on preserving privacy by hiding the power consumption readings. 

\section{Preliminaries} \label{sec:Preliminaries} 
In this section, we present a brief description of the cryptosystems and deep-learning approaches as well as the widely used activation functions that will be used in our scheme.

\subsection {Bilinear Pairing} \label{sub:PPETD_pairing}

Let $\mathbb{G}$ and $\mathbb{G}_{T}$ be two cyclic groups of the same prime order $q$ and $P$ be a generator of group $\mathbb{G}$. Suppose $\mathbb{G}$ and $\mathbb{G}_{T}$ are equipped with a pairing, i.e., a non-degenerated and efficiently computable bilinear map $\hat{e}$: $\mathbb{G} \times \mathbb{G} \rightarrow \mathbb{G}_{T}$ such that:
\begin{itemize}
    \item $\hat{e}(P,P)\neq1_{\mathbb{G}_{T}}$
    \item $\hat{e}\left(a P_{1}, b Q_{1}\right)=\hat{e}\left(P_{1}, Q_{1}\right)^{a b} \in \mathbb{G}_{T}$, for all $a, b \in \mathbb{Z}_{q}^{*}$ and any $P_{1}, Q_{1} \in \mathbb{G}$.
\end{itemize}
We will use bilinear pairing~\cite{pairing} to verify the SMs' signatures efficiently.
 
% We refer to~\cite{boneh2001identity} for a more comprehensive description of pairing technique, and complexity assumptions. 

\begin{figure*}[t]
\centering
\includegraphics[,width=0.85\textwidth]{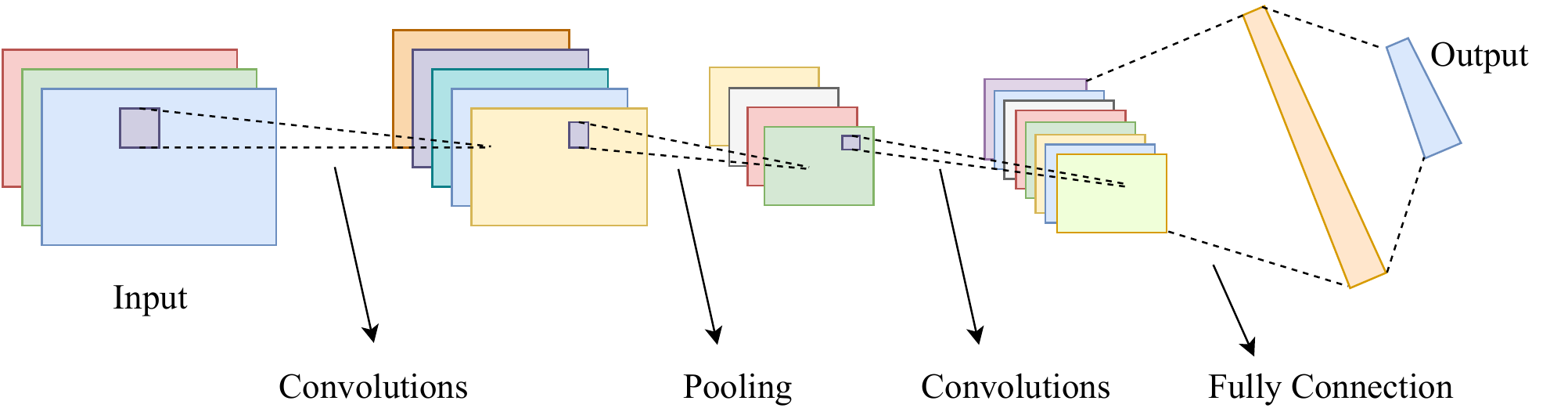}
\caption{Typical architecture of convolutional neural network (CNN).} \label{fig:CCN_arch}
\end{figure*}
%%%%%%%%%%%%%%%%%%%%%%%%%%%%%%%%%%%%%%%%%%%%%%%%%
\subsection{Homomorphic Encryption}
Homomorphic encryption is based on the Paillier cryptosystem that allows certain mathematical operations on the plaintext to be performed directly on the ciphertext~\cite{paillier1999public}. Basically, the Paillier cryptosystem is comprised of three algorithms: key generation, encryption, and decryption.

\begin{itemize}
    \item \textbf{Key generation:}
To construct the public key, two large prime numbers $p_{1}$ and $q_{1}$ are first chosen, where $\left|p_{1}\right|=$ $\left|q_{1}\right|$, and then, $n$ and $\lambda$ are computed as follows: $n=p_{1} q_{1}$ and $\lambda=$ $\operatorname{lcm}\left(p_{1}-1, q_{1}-1\right)$. Next, ${\mu}$ is calculated as follows: ${\mu}=L(g^{\lambda} \bmod n^{2})^{-1}$ mod $n$, where $L(u)=$ $\frac{u-1}{n},$ and a generator $g \in \mathbb{Z}^{*}_{n^{2}}$. Hence, the public key is $p k=(n, g),$ and the corresponding private key is $s k=(\lambda, \mu)$.

\item \textbf{Encryption:}
Let $E(\cdot)$, $m\in\mathbb{Z}_{n},$ and $r\in\mathbb{Z}_{n}^{*}$ be the encryption function, a message, and a random number, respectively. The ciphertext can be calculated as follows.
$$
c=E(m)=g^{m} \cdot r^{n} \bmod n^{2}
$$

\item \textbf{Decryption:}
Given the ciphertext $c\in\mathbb{Z}^{*}_{n^{2}}$, the plaintext is:
$$
m=D(c)=L\left(c^{\lambda \bmod n^{2}}\right) \cdot \mu \bmod n
$$
Then, the additive homomorphic property is as follows.
$$
\begin{aligned}
E\left(m_{1}\right) \cdot E\left(m_{2}\right) &=\left(g^{m_{1}} \cdot r_{1}^{n}\right)\left(g^{m_{2}} \cdot r_{2}^{n}\right) \bmod n^{2} \\
&=g^{m_{1}+m_{2}} \cdot\left(r_{1} r_{2}\right)^{n} \bmod n^{2} \\
&=E\left(m_{1}+m_{2}\right)
\end{aligned}
$$
\end{itemize}
The homomorphic encryption is used in our scheme to aggregate the encrypted power consumption readings of the consumers in an AMI network without revealing the individual readings to preserve privacy.
%%%%%%%%%%%%%%%%%%%%%%%%%%%%%%%%%%%%%%%%%%%%%%%%%%%%%%%%

\subsection{Deep Learning}
Deep learning is a special case of neural networks (NNs) in which multiple hidden layers are used. In general, the structure of NNs consists of three types of layers: input, output, and hidden~\cite{zheng2018wide}. 
% On the other hand, deep networks can extract complex features from unstructured data which is a complex task especially when the attributes are hard to obta in~\cite{7447103}.
There are two types of learning, including supervised and unsupervised learning. In supervised learning, a labeled dataset is used to train a model. Examples of such supervised learning approaches are the multi-layer perceptron (MLP)~\cite{10.5555/1213811}, convolutional neural network (CNN)~\cite{lecun1995convolutional}, and recurrent neural network (RNN)~\cite{ha2018recurrent}. Unsupervised learning, on the other hand, is used when the dataset is unlabeled. Clustering is the widely used approach in unsupervised learning to cluster unlabeled data with similar features together such as K-means clustering algorithm~\cite{clustering}.

The training of a deep-learning model starts with feeding the features/input data into the model layers. Then, these features are gradually mapped into higher-level abstractions via the iterative update (a.k.a, feed-forward and back-propagation) through the intermediate layers of the model for a predefined number of iterations. These mapping abstractions can be used to determine the decision in the output layer.
The model training involves learning the weight and the bias parameters $\Theta$ by defining a cost function and selecting an optimizer.
Using the back-propagation algorithm~\cite{6745416}, the model weights and biases are updated in each iteration using the gradients of the cost function. The output values of the model are compared to the correct values for optimizing the cost function. Then, the error is fed back through the layers to adjust the weights of each connection in order to reduce the cost (loss) function~\cite{6745416}. For the cost function in classification tasks, \textit{categorical cross-entropy} $C(y,\hat{y})$ is defined to measure the loss due to the difference between the true distribution $y$ and the learned distribution $\hat{y}$, for $M$ classes as follows: 

% value to the true distribution  

\begin{equation*}
    \begin{aligned}
        C(y,\hat{y}) = \underset{\Theta}{\min} (-\sum _{c\,=\,1}^{M} y(c)\ log(\hat{y}(c))).\label{eq:cross-entropy}
    \end{aligned}
\end{equation*}

During training, an optimization method, e.g., \textit{Adam}~\cite{kingma2019method}, is used for optimizing the cost function and labeled data are used to train the model. In addition, hyper-parameters of the model such as the number of neurons in each layer, the number of layers, and type of the optimizer, can be adjusted using k-fold cross validation, hyperopt~\cite{Bergstra_2015}, or any other validation methods~\cite{8545748}.

In our scheme, we use CNN and gated recurrent unit neural network (GRU) to train the attacker and defense models, and use K-means clustering to label our dataset.

\subsection{Convolutional Neural Network (CNN)}
CNN is widely used in solving many challenging machine learning problems such as computer vision applications, natural language processing, and speech processing \cite{lecun1995convolutional}. This wide adoption of CNN is due to its capability of capturing complex patterns in the input. 
In the following, we discuss the architecture of a CNN model.

As shown in Fig.~\ref{fig:CCN_arch}, a typical architecture of a CNN model has input, convolution, pooling, fully connected, and output layers. A convolution layer consists of a set of independent filters and max pooling layer(s). The convolution operation is done by sliding each filter, which is usually small in size, over the input to extract features from the input. 
% These filters , and thus, they can interpret input data with strong local connectivity and dependency patterns. 
% Convolving one filter with a channel from the input results in a two-dimensional \textit{feature map} of that filter. 
After the convolution step, a nonlinear activation function such as Rectified Linear Unit (ReLU), Sigmoid, or Tanh is used to enable the model to make complex decisions and solve difficult tasks. 
On the other hand, a pooling layer is used to compress the output of the convolution layer by sub-sampling the feature maps while retaining the most important information. 
Convolution and pooling layers are usually followed by one or more fully connected layers that process the features extracted to be used for inference. 
% We will use CNN in training our models due to its capability of capturing temporal correlation in the input.

%%%%%%%%%%%%%%%%%%%%%%%%%%%%%%%%%%%%%%%%%%%%%%%%%%%%%%%%

\subsection{Gated Recurrent Unit Neural Network (GRU)}
\begin{figure}[t]
\centering
\includegraphics[width=3.1in]{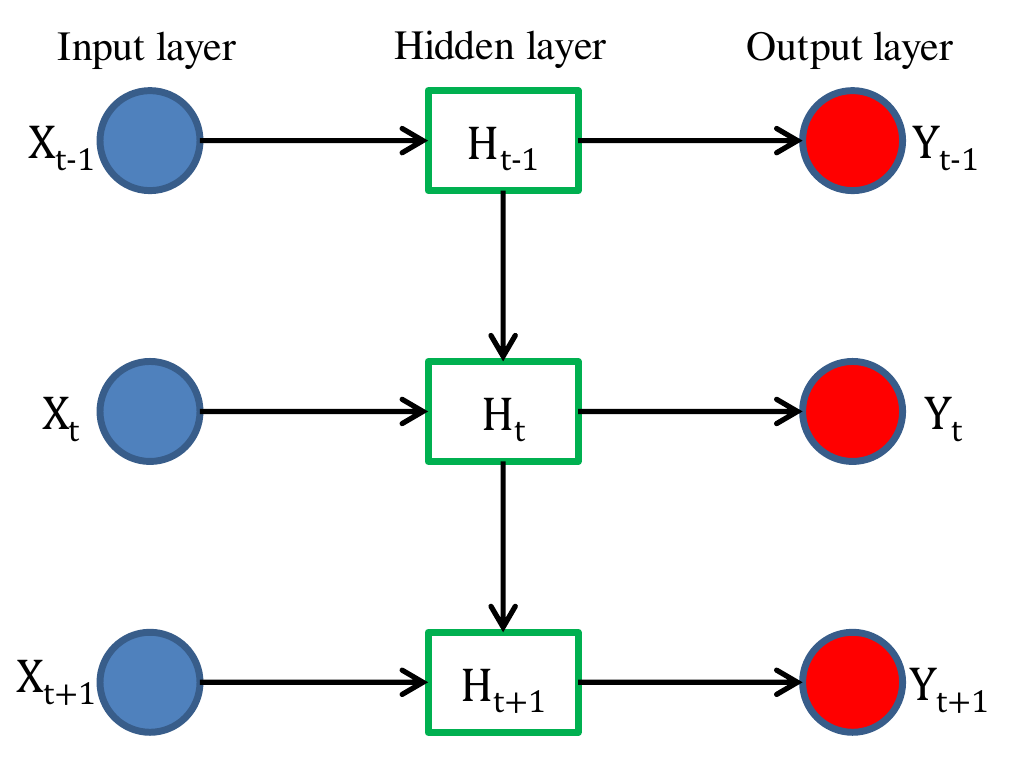}
\caption{Typical architecture of deep recurrent neural network (RNN).} \label{RNN_arch}
\end{figure}

GRU is one type of RNNs that can memorize long sequences of input patterns by using hidden states, sometimes called hidden memory, and building connections between the internal units to form a directed graph as shown in Fig.~\ref{RNN_arch}~\cite{yin2017comparative}. 
The key component in RNN is the transition function in each time step $t$ which takes the current time information $X_{t}$ and the previous hidden state ${H}_{t-1}$ and updates the current hidden state as follows.

\begin{equation}\label{RNN_eq}{H}_{t}={F}\left({X}_{t}, {H}_{t-1}\right),\end{equation}
% \begin{equation}\label{RNN_eq}\mathbf{H}_{t}=\mathbb{F}\left(\mathbf{X}_{t}, \mathbf{H}_{t-1}\right),\end{equation}
where ${F}$ is a nonlinear transformation/activation function, e.g., Sigmoid and Tanh functions. Due to the recurrent structure, ${H}_{t-1}$ in Eq.~\ref{RNN_eq} can be regarded as a memory of previous inputs, i.e., RNN can remember previous inputs in the network's internal state. In GRU, two gates are used, namely reset and update, to control the information flow and learn which information is important to keep and which information can be discarded. 
% which determine whether the hidden state in the previous time step will be discarded and whether the state will be updated, respectively. 
Hence, GRU has the ability to capture the correlations between the input, and that is why we will use it in training our models.
% A memory used in GRU to remember important states in the past.
GRU is usually used in the applications that need to predict the next word of a sentence based on the preceding words such as text generation~\cite{8985885}. It is also commonly used in speech recognition and synthesis~\cite{6638947}.

% At each moment, the GRU receives the current state and the previous state through its update gate, this determines the activation state of its own neurons. At the same time, the reset gate receives both of the above states and determines how much of the input information is to be forgotten. The input at the current moment is then combined with the weight and the output of the reset gate to get the memory content at the current moment through the activation function. Then the update gate receives the memory contents at the current moment and the implicit state from the previous moment to determine the output and implicit state at the current moment~\cite{8851136}.

% Feedback loops are used to process the input data sequence producing the final output, that can also be a data sequence. Such feedback loops allow the retention of information and this effect is often represented as memory that preserves all the measured parameters and links the inputs together and allows sequential and temporal information to be processed by RNNs. The algorithm that is used by the RNNs to loop information back into the network throughout the computational process for updating the weights through the network is back-propagation through time (BPTT)~\cite{37}.

% ANN (artificial neural network) has also developed rapidly and been widely used, and derived convolutional neural network (CNN) with spatial distribution data and recurrent neural network (RNN, recurrent neural network) with  temporal distribution data [21].
%%%%%%%%%%%%%%%%%%%%%%%%%%%%%%%%%%%%%%%%%%%%%%%%%%%%%%%%

\subsection{K-means Clustering Technique}
\begin{figure}[t]
\centering
\includegraphics[width=2.5in]{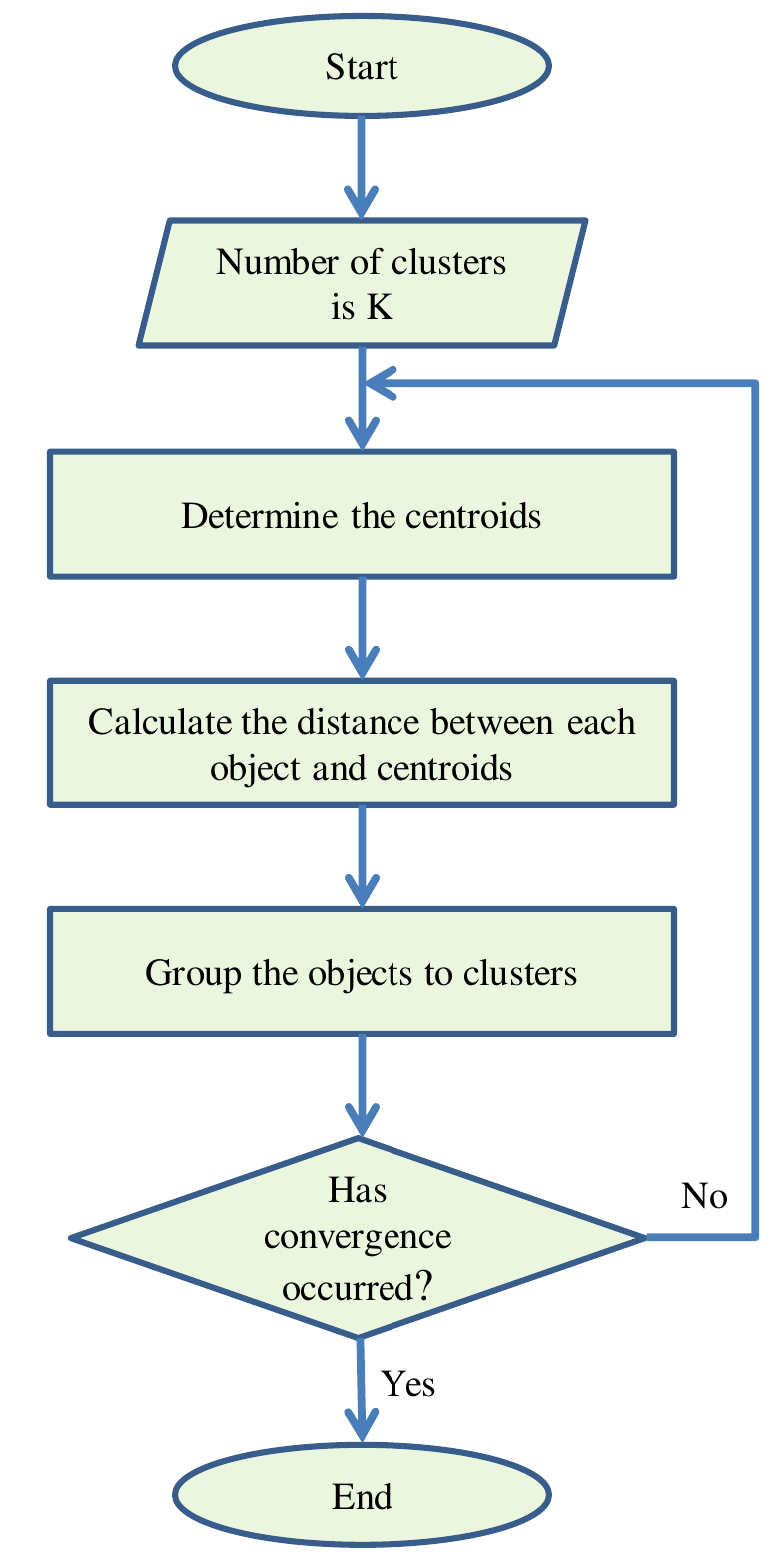}
\caption{The flowchart of K-means clustering technique.} \label{clustering}
\end{figure}

The K-means clustering technique is an unsupervised learning algorithm that is used to classify a given unlabeled dataset, which contains data objects/points, into a certain number of clusters based on the minimum distance between the cluster center and the object~\cite{clustering,dash2010hybridized}. A cluster is a collection of data objects that are homogeneous within the same cluster and heterogeneous to the objects in the other cluster(s)~\cite{dash2010hybridized}.
The main idea is to define $k$ centroids, one for each cluster, and each point in the dataset needs to be associated to the nearest centroid.
Then, $k$ new centroids need to be re-calculated, and thus, new bindings have to be done between the same dataset points and the nearest new centroids. This process should be repeated several times, and in each time, the centriods' locations are changed until convergence occurs when the centroids' locations are not changed any more. 
This iterative process can be summarized in Fig.~\ref{clustering}.
Basically, the clustering algorithm aims at minimizing an objective function, e.g., a squared error function, as follows.
$$
\mathrm{J}=\sum_{j=1}^{\mathrm{k}}
\sum_{i=1}^{n}\left\|\mathrm{x}_{i}-\mathrm{c}_{\mathrm{j}}\right\|^{2},
$$
where $\left\|\mathrm{x}_{i}-\mathrm{c}_{\mathrm{j}}\right\|^{2}$ is the Euclidean distance between a data point $\mathrm{x}_{i}$ and the cluster center $\mathrm{c}_{\mathrm{j}}$, for a given dataset of $n$ data points.

% \textbf{K-Means Clustering Algorithm: }
% The K-means clustering algorithm is composed of the following steps.
% \begin{enumerate}
%     \item Determine the number of clusters
%     \item Place $k$ points into the space represented by the objects that are being clustered. These points represent initial group centroids.
%     \item Calculate the distance of each object to the centroid and assign each object to the group that has the closest centroid. Euclidean distance is generally considered to determine the distance between each data object and the cluster centers~\cite{5453745}.
%     \item When all objects have been assigned, recalculate the positions of the $k$ centroids by computing the average of the early formed clusters.
%     \item Repeat Steps 3 and 4 until the centroids no longer move.
% \end{enumerate}

We will use K-means clustering technique to label the consumers' records as ``absent'' and ``present'', to be used in training our models.
%%%%%%%%%%%%%%%%%%%%%%%%%%%%%%%%%%%%%%%%%%%%%

\subsection{Activation Functions}
The activation functions are used in the machine learning models to transform the summed weighted input from a neuron into the activation of that neuron. They have a major impact on the model accuracy and convergence speed. In the following, we explain the widely used activation functions and their usage. 

\begin{itemize}
    \item \textit{Rectified Linear Unit} (ReLU): It outputs the same input if the input is a positive value, otherwise, the output is zero~\cite{nwankpa1811activation}. ReLU is computationally efficient since it only requires a simple max() function as follows. 
    \begin{align*}
        ReLU(x) = max(0,x).
    \end{align*} 

    % ReLU acts like a linear activation function, and it is usually easier to optimize the model when its behavior is linear or close to linear.

    \item \textit{Softmax}: It is commonly used in the output layer for multi-class classification problems.  For a given input vector, Softmax calculates a probability vector, i.e., for an input vector $\mathbf {z}=[\mathbf {z}[1],\ldots ,\mathbf {z}[M]]\in \mathbb {R} ^{M}$ of length $M$, where $M$ is the number of classes, the Softmax function is defined as follows~\cite{nwankpa1811activation}.

\begin{equation*}
    Softmax ({z}[i])={\frac {e^{z[i]}}{\sum _{j=1}^{M}e^{z[j]}}} \  \ for \ \ i = \{1, \dots, M\}.
     \label{eq:1}
\end{equation*}

% \item \textit{Exponential Linear Unit} (Elu):

\end{itemize}

\section{Dataset Preparation}\label{dataset_preprocessed}

% \textbf{Dataset labeling:}
In this paper, we used a real smart meter dataset, which is available online and released by the Smart project~\cite{PCRdataset}. The dataset contains real power consumption readings of $114$ single-family apartments over $349$ days during the year $2016$ from January $1^{st}$ to December $14^{th}$ at one-minute granularity. %, and thus more bandwidth is consumed. 
Hence, the total number of records is $39,786$, where each record corresponds to power consumption readings of one consumer in a single day. The dataset is then divided into two sets, 80\% for training and 20\% for testing.
From the one-minute granularity (one reading/minute) dataset, we have created other datasets with different transmission rates, including a reading every 5 minutes, 15 minutes, and 30 minutes, by aggregating the power consumption readings. Note that we use ``1/$x$ min'' to refer to transmission rate, which means
one reading is transmitted every $x$ minutes, where $x$ is the transmission interval.
% based on the transmission rate value, e.g., aggregating the last 5 readings every 5 minutes if a reading is transmitted every 5 minutes. 
Using these datasets with thresholds of $ 1\%$, $ 4\%$, $ 7\%$, and $ 10\%$, we create datasets for CAT approach, in which an SM reports its power consumption reading if the absolute value of the percentage of the change in the consumption compared to the last reported reading exceeds the threshold.

Table~\ref{efficiency} gives the efficiency that can be achieved by using CAT approach in terms of the percentage of readings that are not transmitted compared to periodic transmission of readings at different thresholds and transmission rates. We observe that as the threshold increases, the efficiency increases since the probability that the absolute value of the percentage of the change in the power consumption exceeds the threshold decreases, and hence, there are fewer transmissions. 
On the other hand, as the transmission interval increases, the efficiency decreases because the probability  that the absolute value of the percentage of the change in the power consumption exceeds the threshold increases, and hence, there are more transmissions.

\begin{figure*}[t]
    \centering
    \subfloat[Using $ 5\%$ threshold.]{\label{CDF_5}\includegraphics[width=0.5\textwidth]{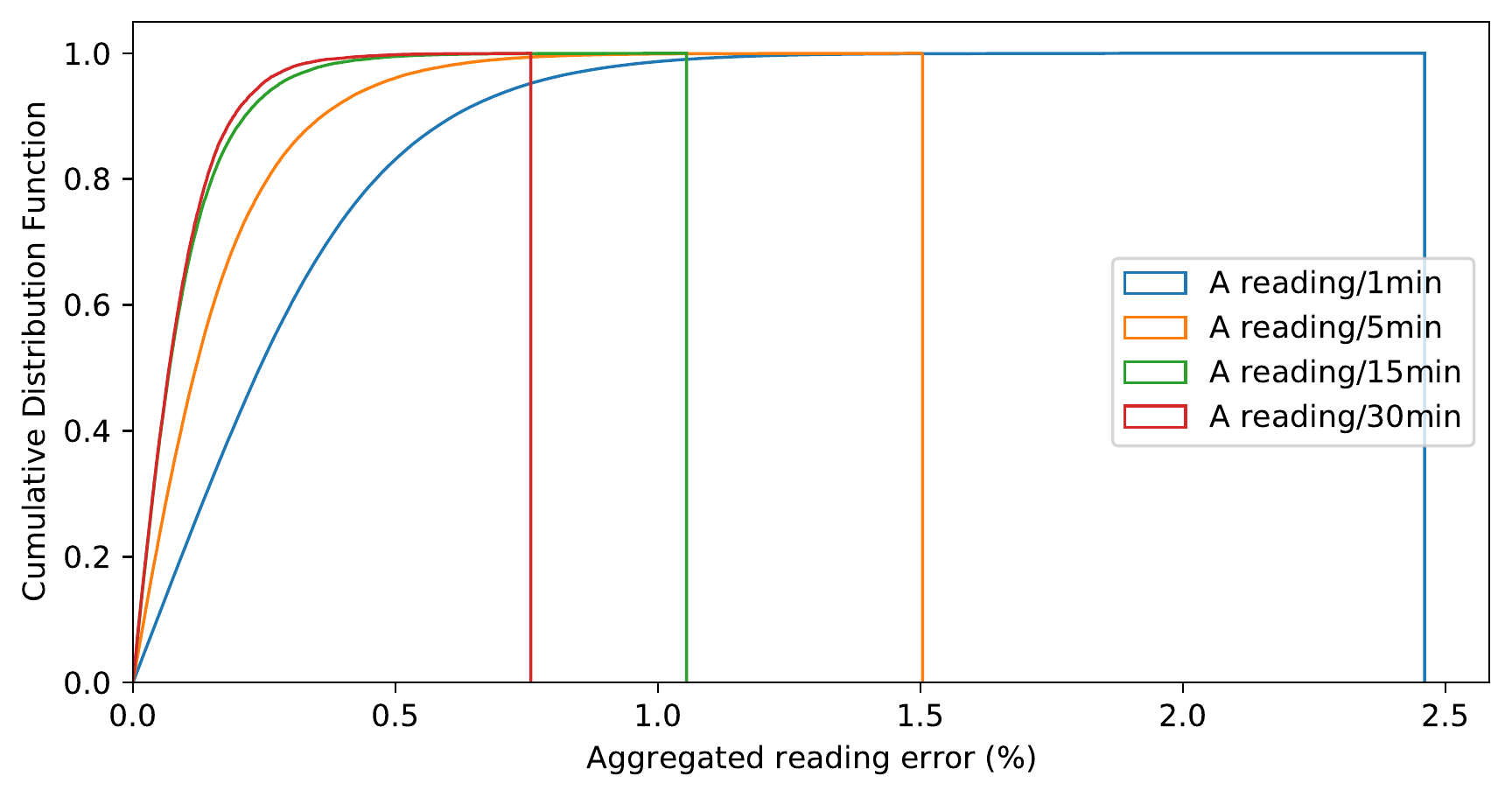}}
    \subfloat[Using $ 10\%$ threshold.]{\label{CDF_10}\includegraphics[width=0.5\textwidth]{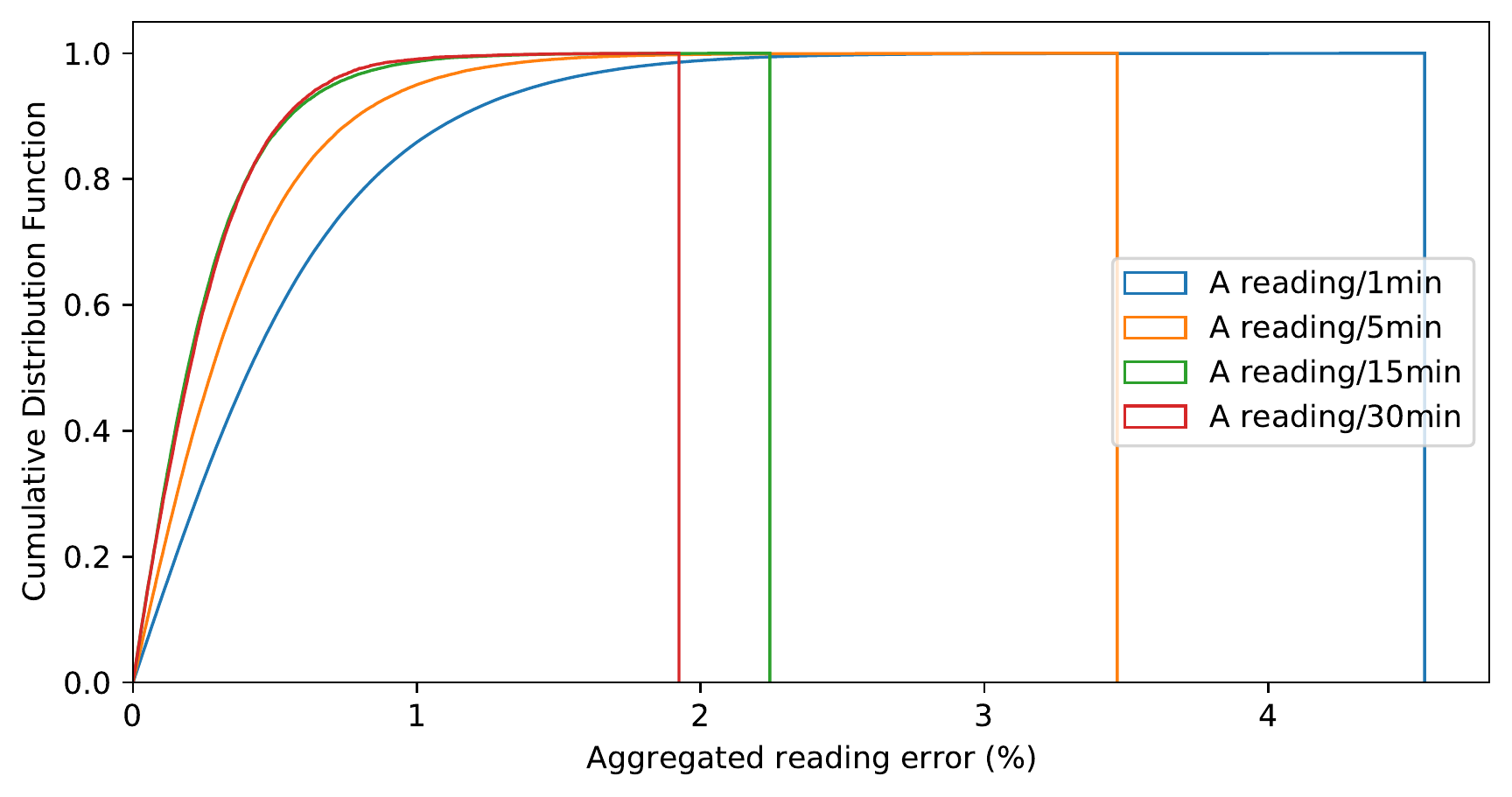}}
    \centering
    \caption{Comparison of CDF of aggregated reading error for different transmission rates at $ 5\%$ and $ 10\%$ thresholds. \label{CDF}}
\end{figure*}

Using CAT approach, there may be a reading error because the reading that is considered by the EU may be less/more than the actual reading that is measured by the SM due to using a threshold. Therefore, the reading error at time $t$ is ($m_{SM}[t]-m_{EU}[t]$), where $m_{SM}[t]$ is the actual power consumption reading measured by the SM at time $t$ and $m_{EU}[t]$ is the reading that is used by the EU.
Since the EU receives the aggregated reading of a number of SMs, we measured the error in the aggregated reading using our datasets assuming that the AMI network has $114$ SMs. Then, we plot, in Fig.~\ref{CDF}, the cumulative distribution function (CDF) of the aggregated reading error at $5\%$ and $10\%$ thresholds using different transmission rates. 
It can be seen that the aggregated reading error in the worst case does not exceed half of the threshold, which indicates that the error in the aggregated reading is much smaller than the error of the individual readings because the positive errors in the individual readings counter the negative errors, which can reduce the aggregated reading error. 
Moreover, we observe that as the transmission interval increases, the error of the aggregated reading decreases because the probability that the absolute value of the percentage of the change in the power consumption exceeds the threshold increases, and hence, there are more transmissions which results in smaller error of the individual readings.
% most of the readings received by the EU are the real readings.

{\renewcommand{\arraystretch}{1.35}
\begin{table}[t]
\centering
\caption{Efficiency (\%) achieved by using CAT approach at different thresholds and transmission rates.}
\label{efficiency}
\resizebox{1\columnwidth}{!}{%}
\begin{tabular}{>{\centering\arraybackslash}m{1.5cm} >{\centering\arraybackslash}m{1cm}|
>{\centering\arraybackslash}m{1cm}|
>{\centering\arraybackslash}m{1cm}|
>{\centering\arraybackslash}m{1cm}|
>{\centering\arraybackslash}m{1cm}|}
\hhline{~~|----}
 &  & \multicolumn{4}{c|}{\textbf{\cellcolor[gray]{0.8}Threshold (\%)}} \\ \cline{3-6} &&  $ 1$ & $4$&$ 7$&$ 10$       \\ \hline
\multicolumn{1}{|c|}{\cellcolor[gray]{0.8}}   & 1/min &    59.69    &     72.19   &   75.81  &    77.96\\ \cline{2-6} 
\multicolumn{1}{|c|}{\cellcolor[gray]{0.8}}                       &  1/5min & 19.99&  31.72  &     36.48 & 39.75\\ \cline{2-6} 
\multicolumn{1}{|c|}{\cellcolor[gray]{0.8}}                       &  1/15min  &6.91 &  13.91  & 18.24  &21.64 \\ \cline{2-6} 
\multicolumn{1}{|c|}{\multirow{-4}{*}{\cellcolor[gray]{0.8}\begin{tabular}[c]{@{}c@{}}\textbf{Transmission} \\ \textbf{rate}\\\end{tabular}}} & 1/30min &4.75 & 11.92& 17.17 & 21.77\\ \hline
\end{tabular}}
\end{table}}

\begin{figure*}[t]
    \centering
\subfloat[Consumption pattern of consumer 1.]{\label{fig:consp1}\includegraphics[width=2.5in]{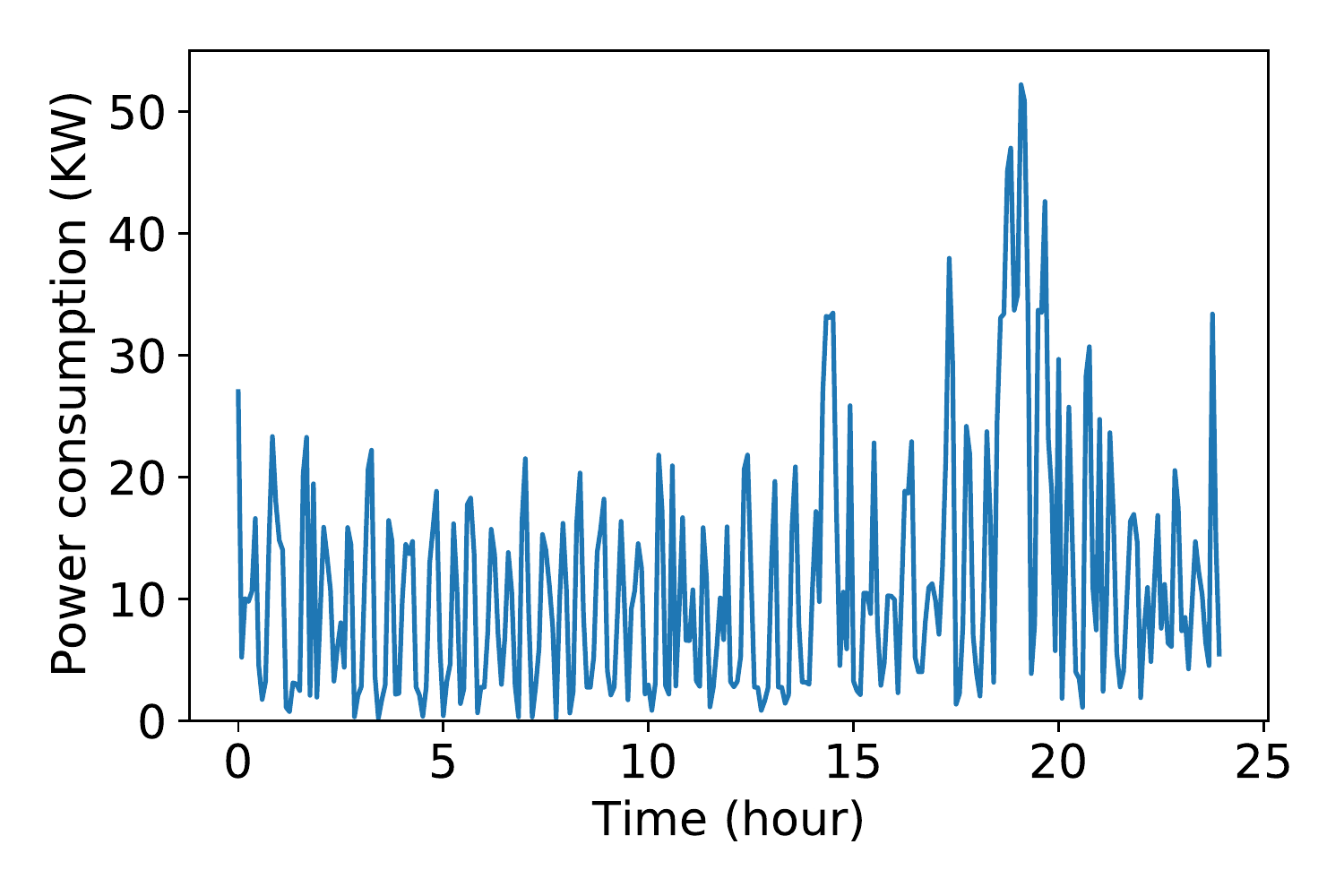}}
\subfloat[Consumption pattern of consumer 2.]{\label{fig:consp2}\includegraphics[width=2.5in]{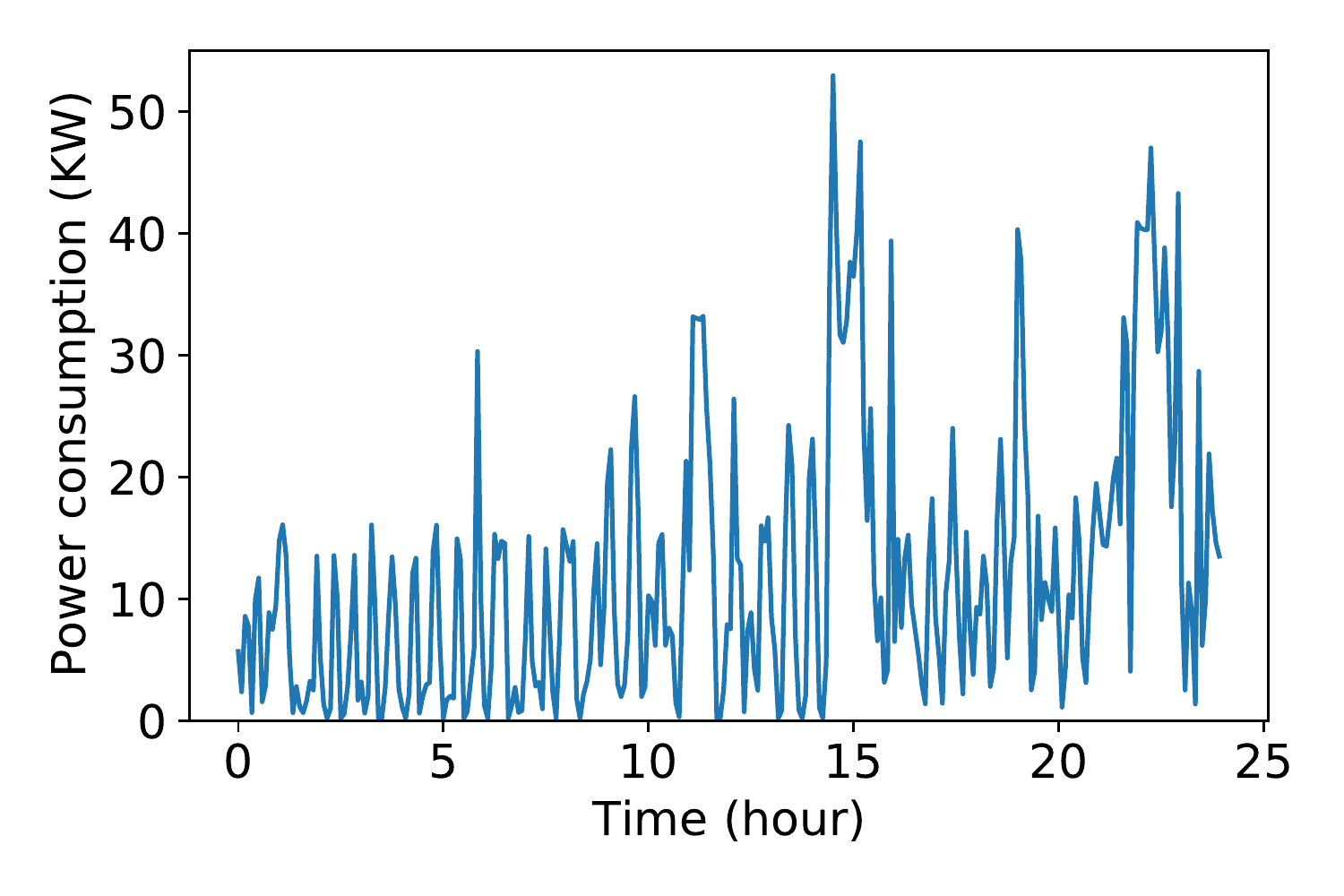}}
\subfloat[Consumption pattern of consumer 3.]{\label{fig:consp3}\includegraphics[width=2.5in]{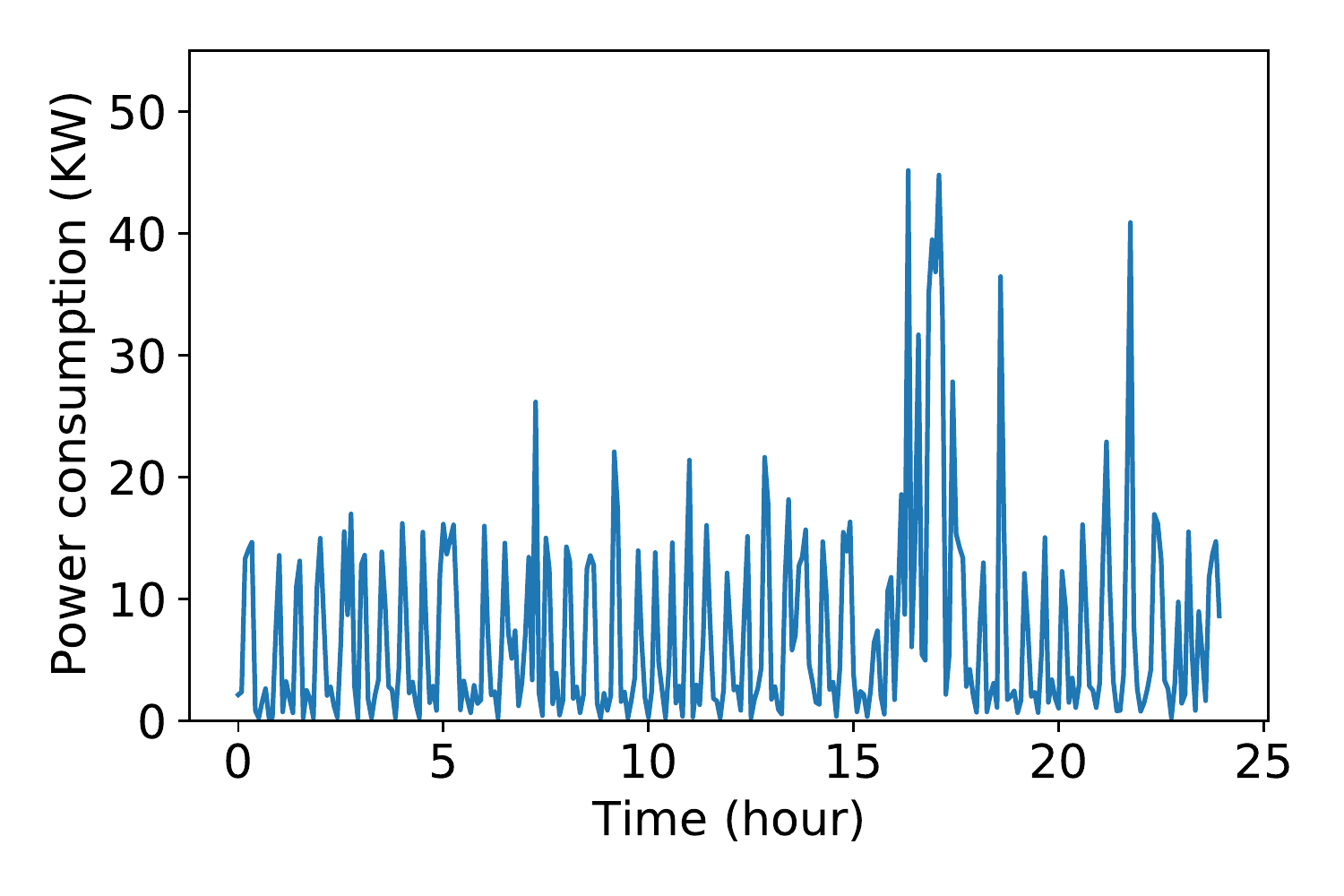}}
\hfil

\subfloat[Transmission pattern of consumer 1.]{\label{fig:transp1}\includegraphics[width=2.5in]{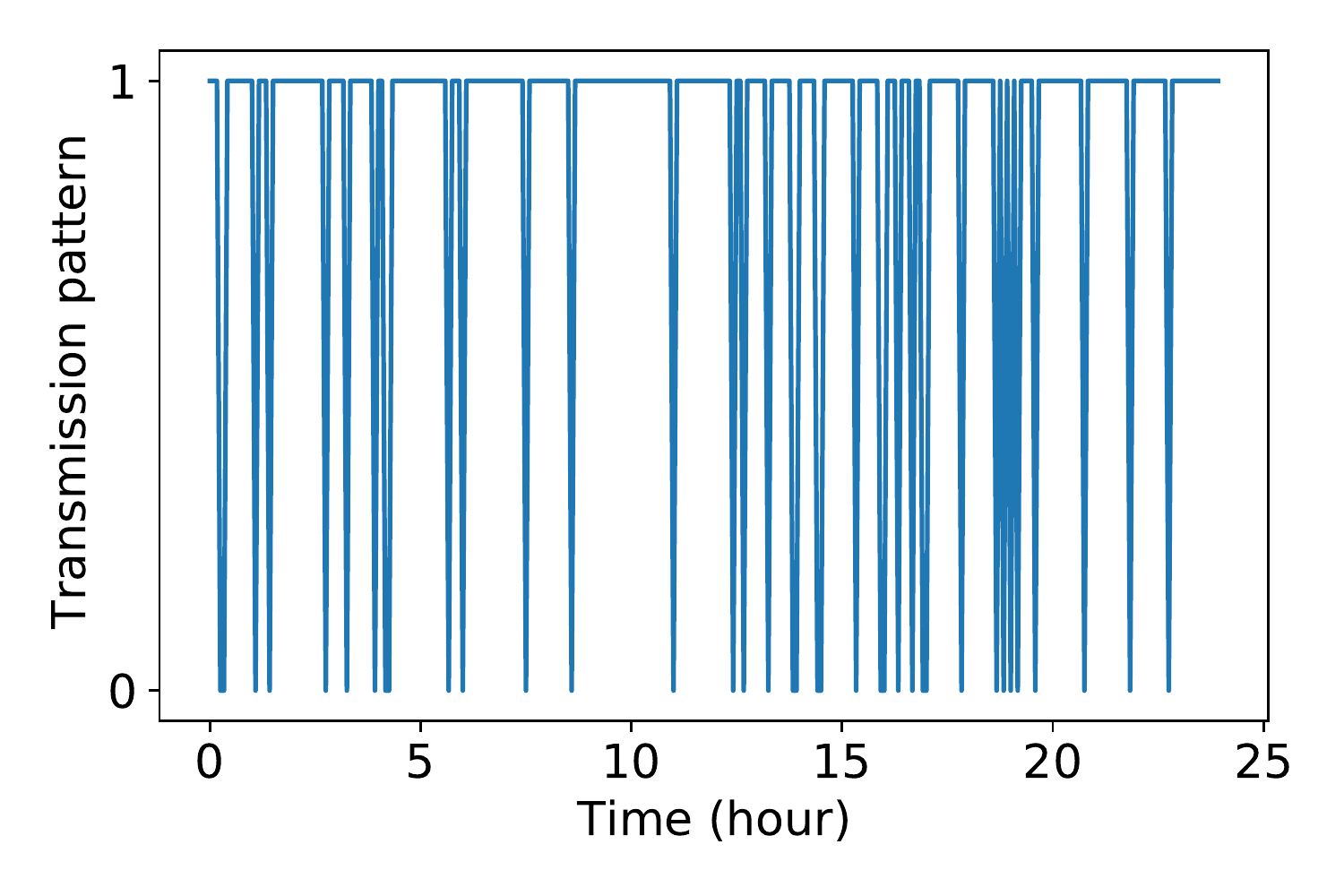}}
\subfloat[Transmission pattern of consumer 2.]{\label{fig:transp2}\includegraphics[width=2.5in]{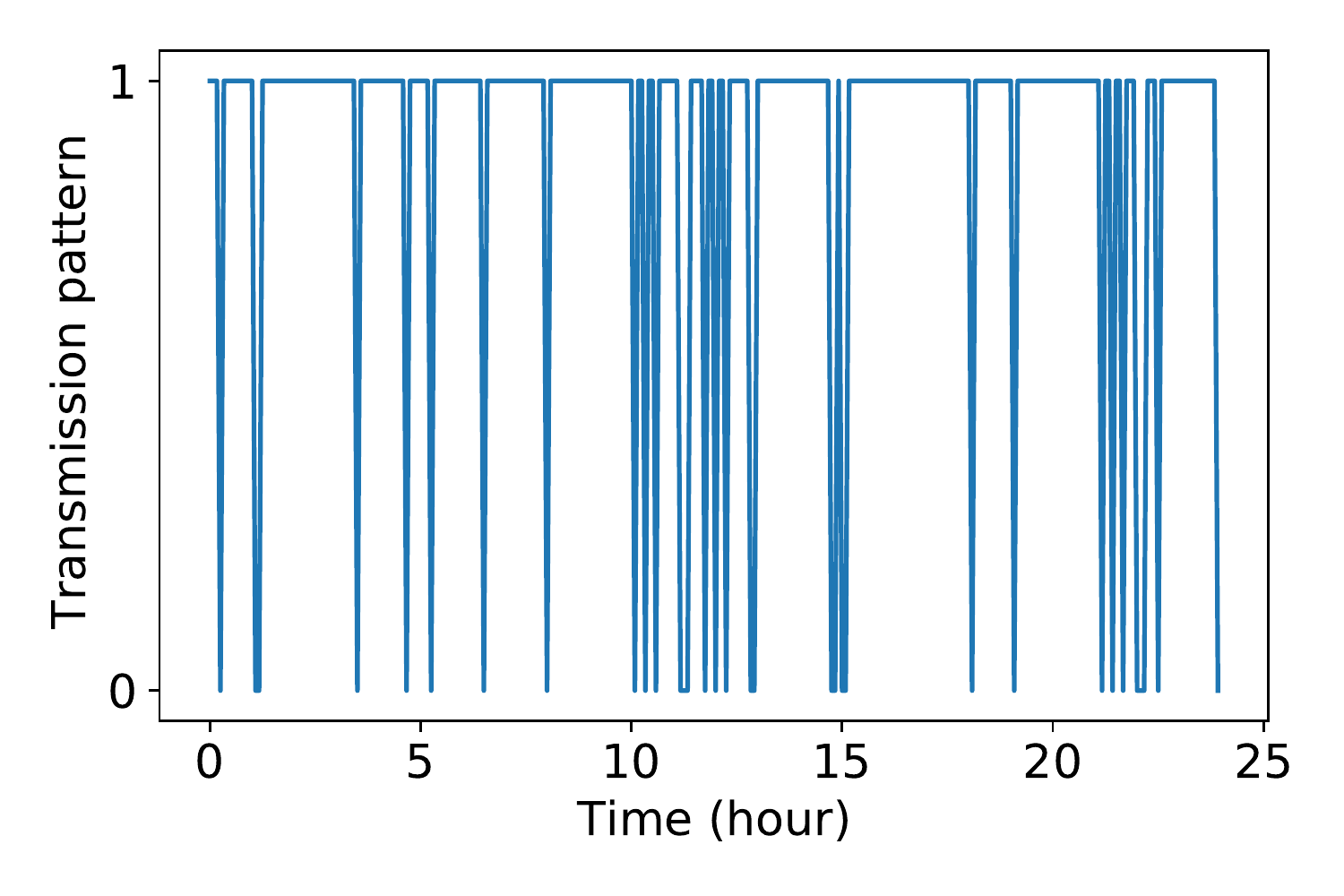}}
\subfloat[Transmission pattern of consumer 3.]{\label{fig:transp3}\includegraphics[width=2.5in]{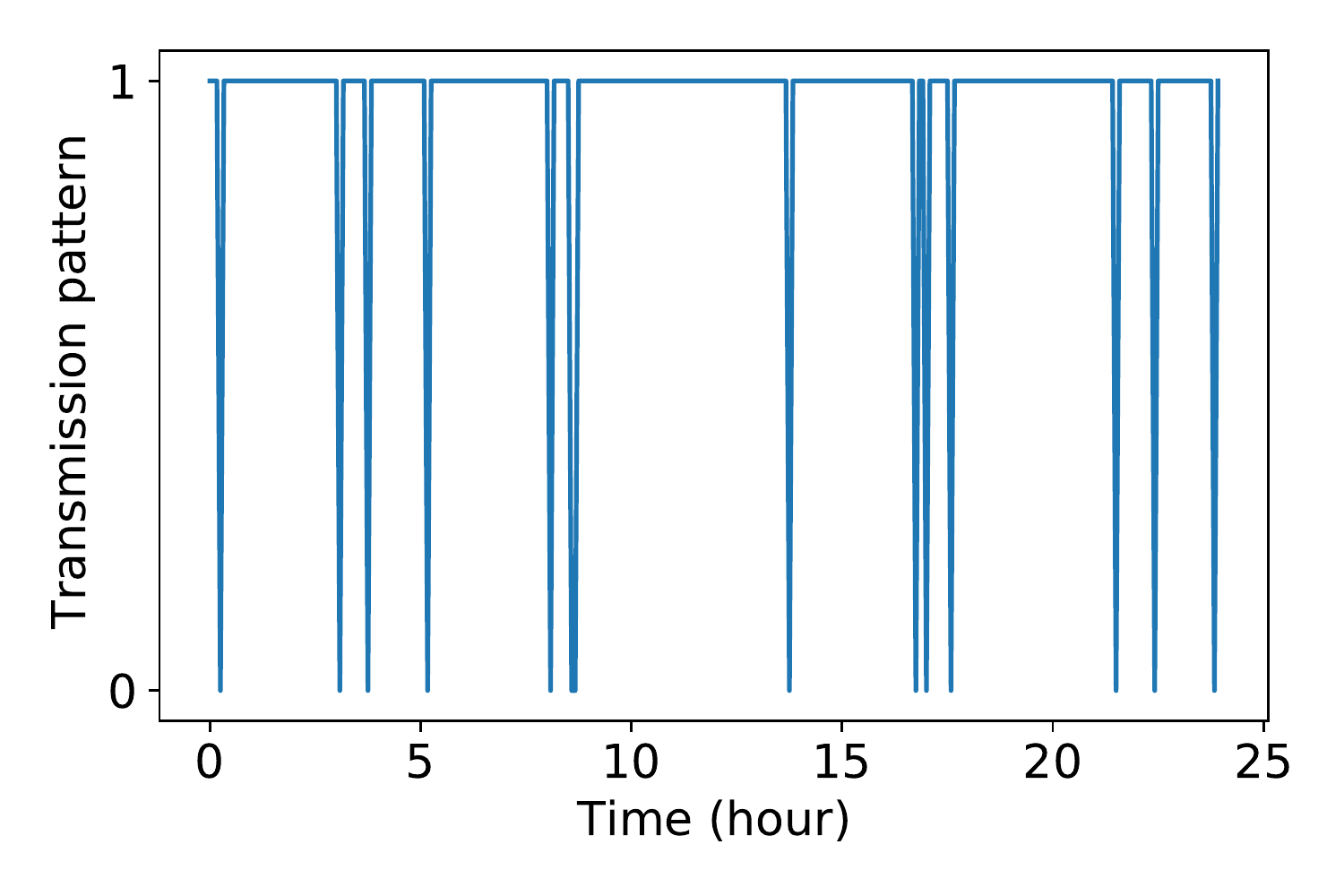}}
    \centering
    \caption{Consumption and transmission patterns of different consumers when they are present at home, where in (d)-(f), 1 and 0 refer to transmission and no transmission events, respectively. \label{present}}
\end{figure*}

\begin{figure*}[t]
    \centering
\subfloat[Consumption pattern of consumer 1.]{\label{fig:consa1}\includegraphics[width=2.5in]{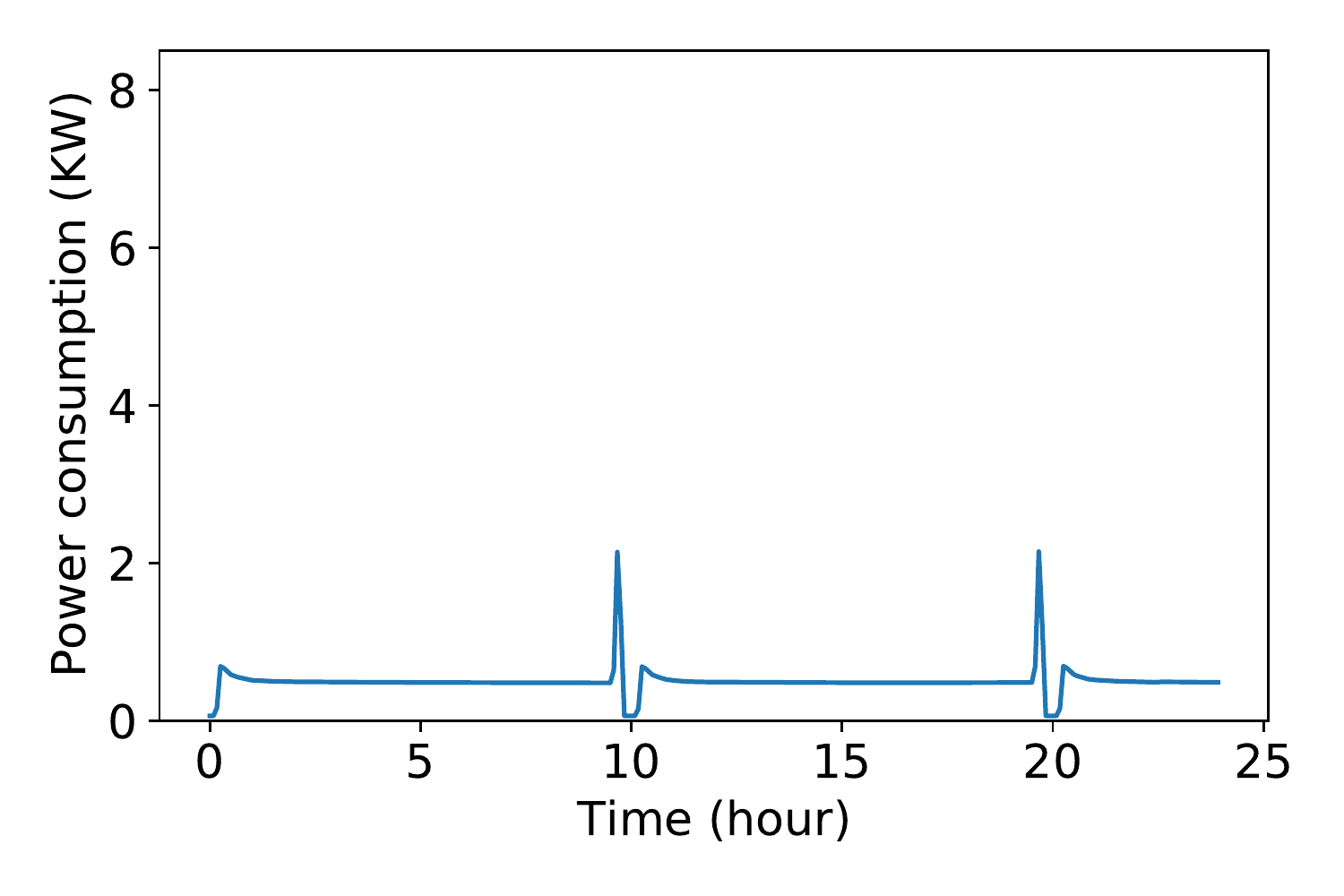}}
\subfloat[Consumption pattern of consumer 2.]{\label{fig:consa2}\includegraphics[width=2.5in]{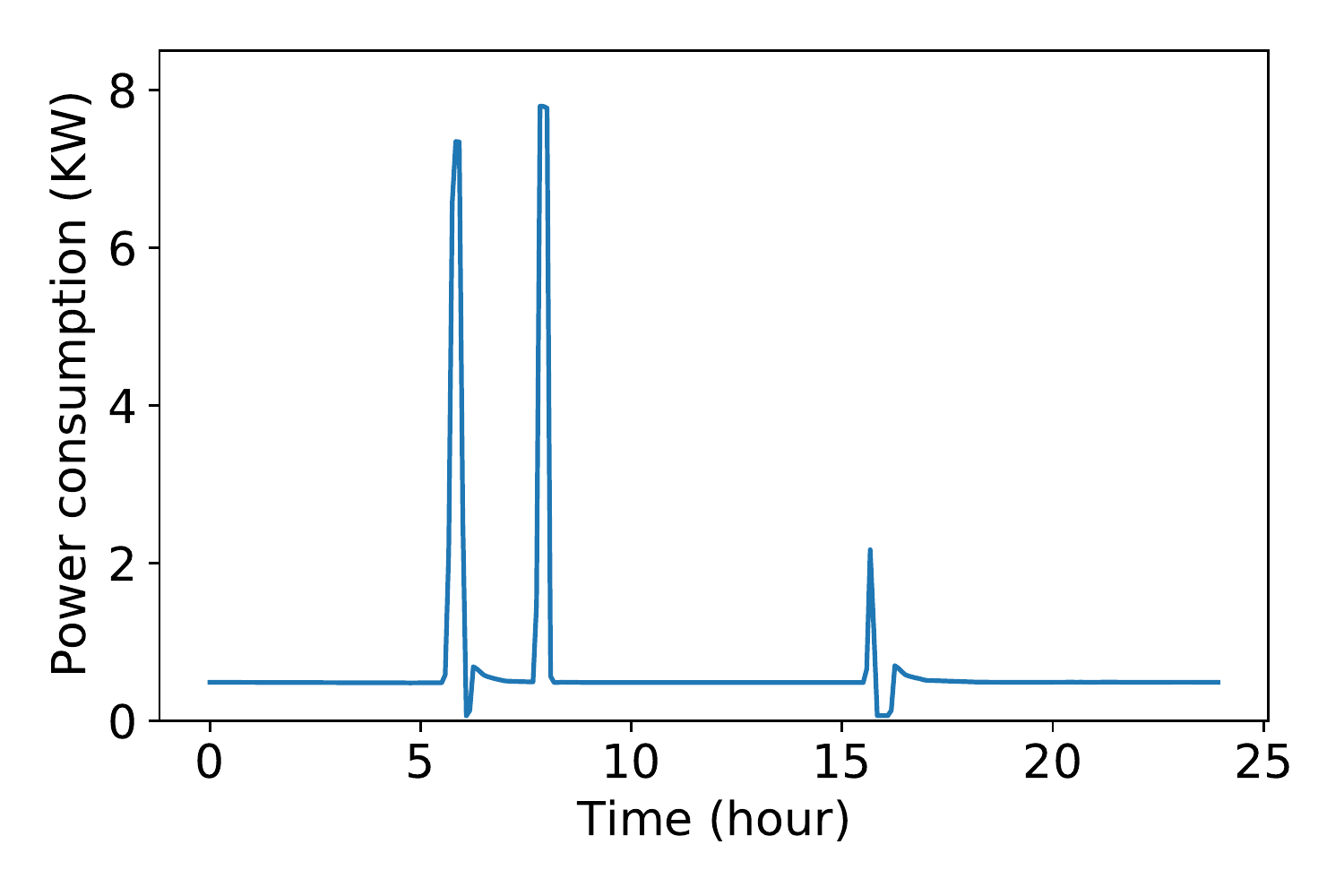}}
\subfloat[Consumption pattern of consumer 3.]{\label{fig:consa3}\includegraphics[width=2.5in]{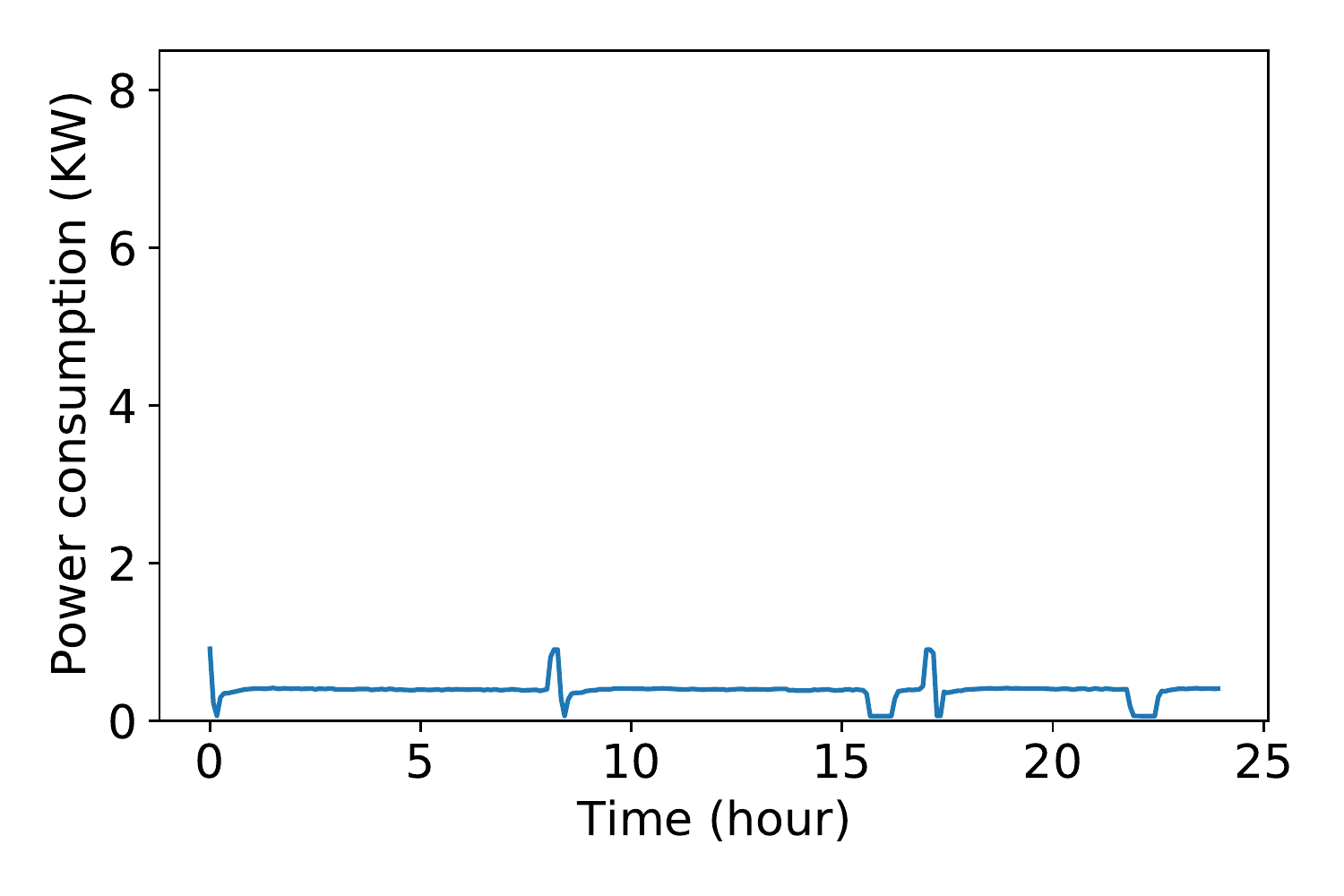}}
\hfil

\subfloat[Transmission pattern of consumer 1.]{\label{fig:transa1}\includegraphics[width=2.5in]{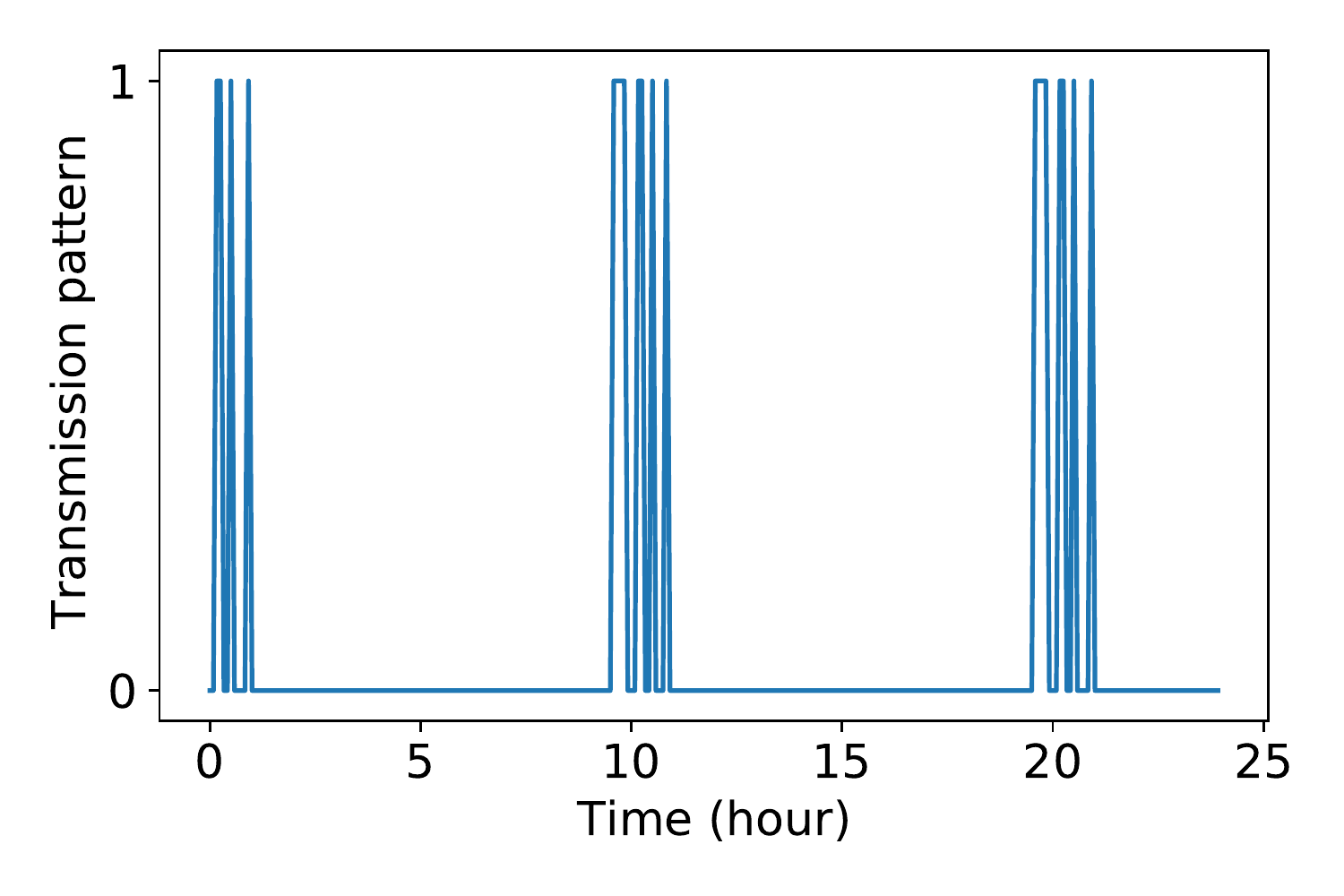}}
\subfloat[Transmission pattern of consumer 2.]{\label{fig:transa2}\includegraphics[width=2.5in]{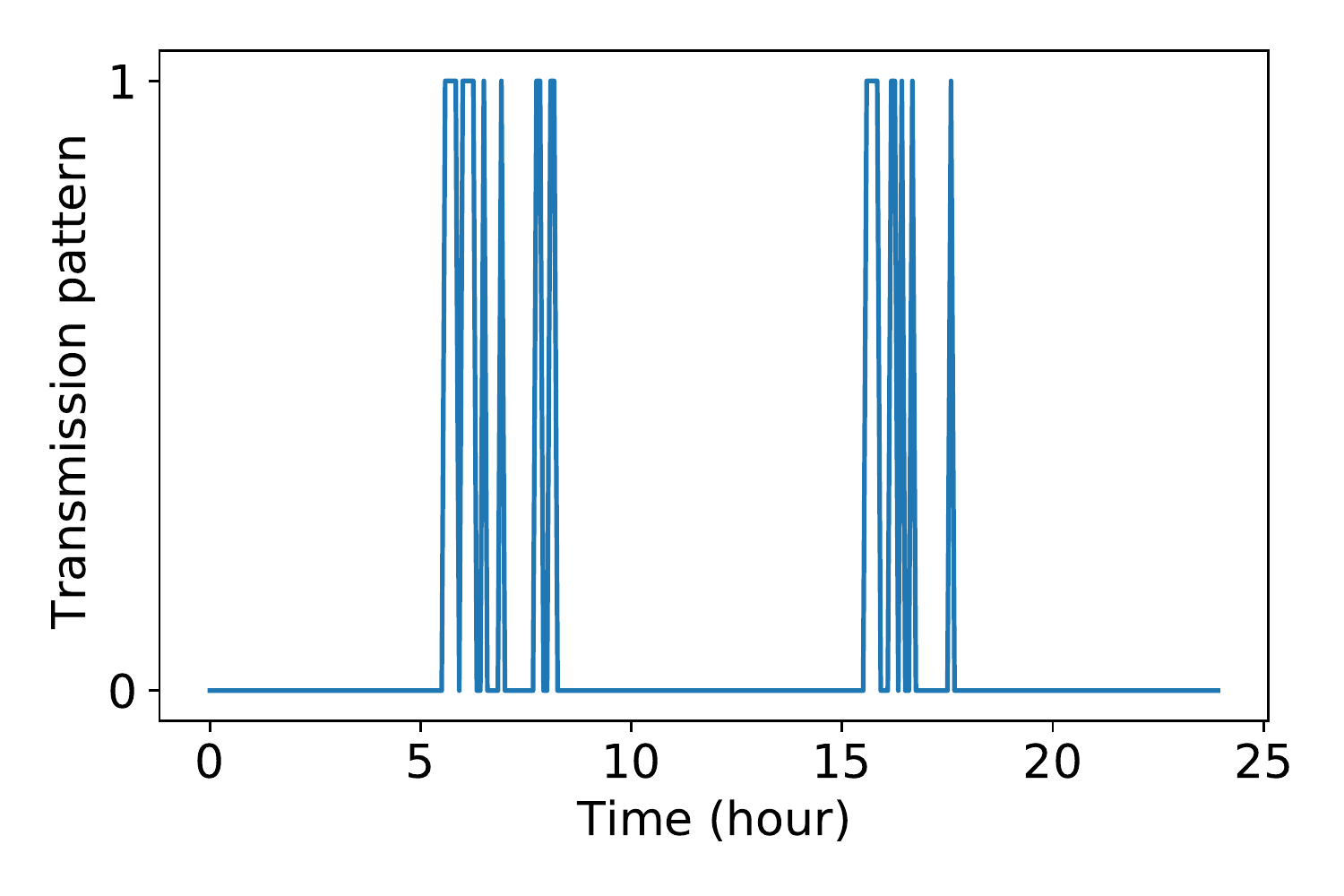}}
\subfloat[Transmission pattern of consumer 3.]{\label{fig:transa3}\includegraphics[width=2.5in]{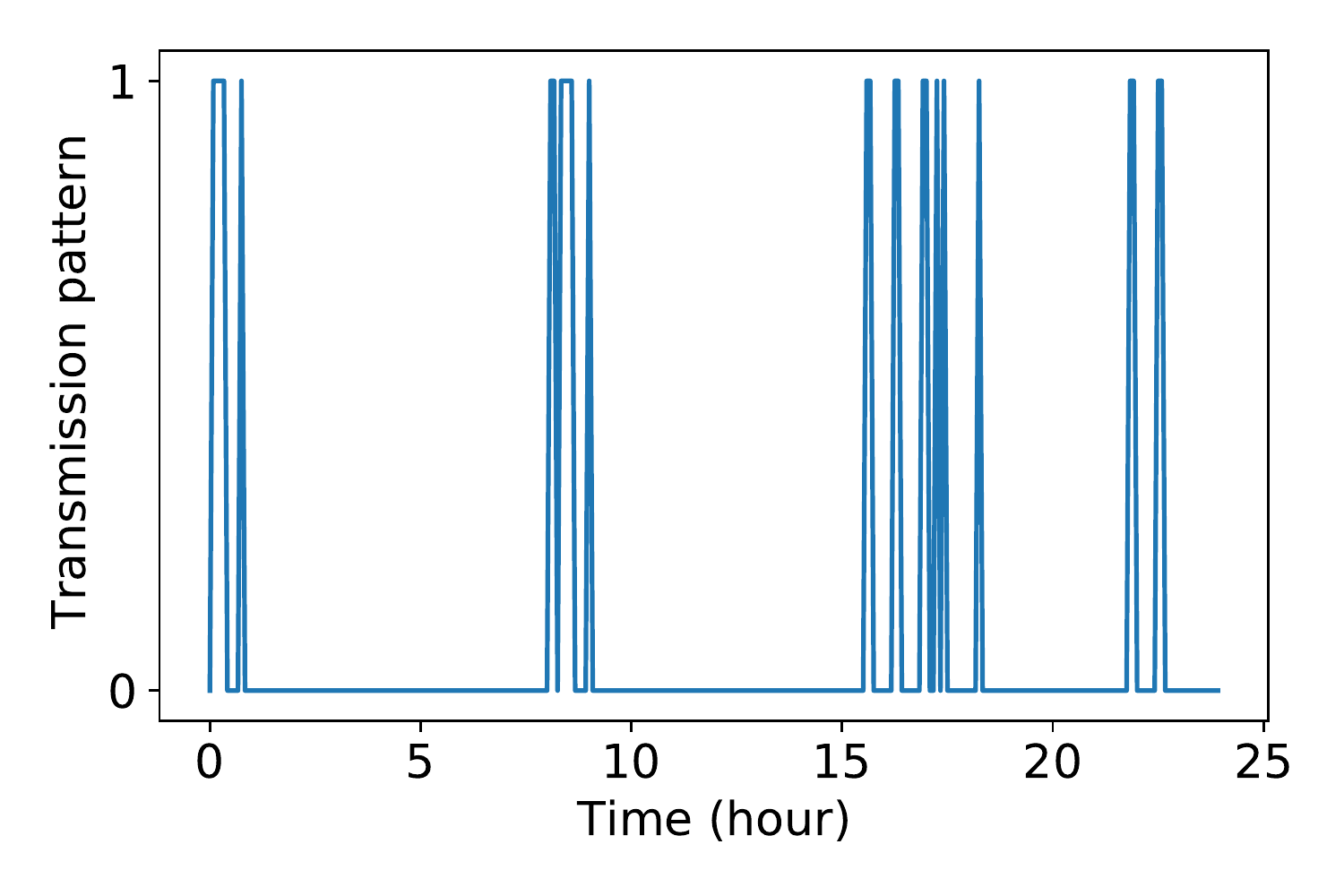}}
    \centering
    \caption{Consumption and transmission patterns of different consumers when they are not at home, where in (d)-(f), 1 and 0 refer to transmission and no transmission events, respectively. \label{absent}}
\end{figure*}

% Compared to the CAT approach that proposed by~\cite{Dr_Li} in which the measurement is transmitted only if the difference between two consecutive measurements exceeds the threshold, we compared the current power consumption reading with the last reported reading.

On the other hand, the existing power consumption datasets do not include the presence and absence of the consumers. Hence, in this section, we aim at labelling the consumers' records per day as absent or present, to be used in training our models, by using the following two approaches.

\begin{enumerate}
    \item \textbf{Clustering:}
        We used K-means clustering algorithm~\cite{clustering} for each consumer's record to label it absent or present.
        % to label/classify them as ``present'' when the house dwellers are at home or ``absent'' when they are not at home. 
        % Generally, K-means algorithm is a clustering algorithm which aims to divide the dataset into k clusters/classes. 
        The intuition behind using K-means clustering is that when the dwellers are not at home, their transmission pattern differs from the pattern when they are at home. So a clustering approach can classify the dataset records into two classes; one for ``present'' when the house dwellers are at home and ``absent'' when they are not at home.
        We clustered each consumer separately since each home has its own transmission pattern that depends on the lifestyle of its dwellers, the number of dwellers, etc.
        % On the other hand, if the clustering is done to all the consumers as a single consumer, it may consider a present high usage of a certain consumer with an absent low usage of another consumer.
        
        % After applying the clustering algorithm, we took the least centroid cluster that represents the least power consumption as the absent class.

        % This is done by finding the optimal number of clusters that provides about 20\% of the total consumer's days are on-travel.

    % \item \textbf{Criteria 2: ``Monthly average''}
    %     For each consumer, We considered the total consumption in each day and compare it with the consumer's monthly average by using the following equation for each day for each consumer. Monthly average is expected to be a suitable criteria since the power consumption pattern in summer is different from winter.
        
    %     \[ C_2
    %     = \dfrac{T_{cdm}-avg_{cm}}{avg_{cm}},
    %     \]
    %     where $T_{cdm}$ is the total consumption for consumer $c$ on day $d$ in month $m$, while $avg_{cm}$ is the monthly average power consumption of consumer $c$ in month $m$. Then, the lowest 20\% of the total resulted records are considered as on-travel.
    
    \item \textbf{Periods' comparison:}
        In this approach, we divide the day into three periods, $t_1, t_2, $ and $t_3$, that correspond to the following time periods, 8AM-4PM, 4PM-12AM, and 12AM-8AM, respectively. Then, if a consumer is present at home, the consumption at period $t_2$ is usually different from the consumption at this period because the consumer does many activities at the other periods such as cooking, as confirmed by~\cite{Dr_Li}.
        First, we calculate the total consumption of each period $t_i$, where $i=\{1,2,3\}$, and then, we compute the following formula for each consumer's record that has daily power consumption readings.
         \[ \left|\dfrac{C_{1}[d]-C_{2}[d]}{C_{1}[d]}\right| + \left|\dfrac{C_{3}[d]-C_{2}[d]}{C_{3}[d]}\right|,
        \]
        where $C_{i}[d]$ is the total power consumption of the $i^{th}$ period in day $d$. The result of this formula is low when the consumption at periods $t_1$ and $t_3$ are close to the consumption at period $t_2$, which can indicate that the consumer is absent.
        
        % Using this formula, we can know how far the consumption at period $T_2$ from periods $T_1$ and $T_3$ in such a way the consumer is absent as the difference gets lower.
        % Furthermore, Fig.~\ref{} shows the comparison between the power consumption changes in the midnight and those during the cooking time. We observe a substantial difference.
        
\end{enumerate}

Finally, the records in our datasets are labelled absent if both approaches label them absent, otherwise they are labelled present.  

In this paper, we study two use cases at $ 10\%$ threshold; one using 1/5min (transmitting a reading every 5 minutes) and another using 1/30min (transmitting a reading every 30 minutes) transmission rates.
Based on this threshold and transmission rates, we converted the datasets that have power consumption readings into datasets that have transmission patterns, containing zero or one in each time slot indicating no transmission event (when the absolute value of the percentage of the change in the consumption is below the threshold) or transmission event (when the absolute value of the percentage of the change in the consumption is above the threshold), respectively. 
Therefore, we have a transmission dataset that contains consumers' records, where each record is labelled \textit{absent} or \textit{present}, and has $288$ and $48$ binary features, for 1/5min and 1/30min transmission rates, respectively. 
Fig.~\ref{present} and Fig.~\ref{absent} show the power consumption and transmission patterns for three different consumers in our datasets when they are present at home and absent, respectively. As can be seen in the figures, there is a substantial difference in the transmission patterns based on the presence/absence of the consumers.
Attackers can exploit this difference to use the transmission patterns to learn whether a consumer is present or absent. 
In next section, we will use this dataset for training and evaluating both the attacker and defense models.

\section{Proposed scheme}
\label{PS}

In this section, we first use the dataset created in Section~\ref{dataset_preprocessed} to train an attacker model to launch PPA. The model takes the transmission pattern of a consumer as input and detects if the consumer is absent or present. Then, we train a deep-learning-based defense model that can generate spoofing transmissions (redundant real readings) to thwart PPA. Finally, we will discuss our privacy-preserving reading collection scheme.

\subsection{Attacker Model}\label{att_model}

As explained in Section~\ref{dataset_preprocessed}, when a consumer is present at home, the transmission pattern is distinguishable due to using appliances comparing to the transmission pattern when the consumer is absent.
In this subsection, we explain how attackers can make use of this fact by training a deep-learning model to launch PPA to learn whether a consumer is on travel (absent) or at home (present). 
% The attacker uses old absent and present records of the consumers' transmission patterns to train a deep learning model that is used to launch PPA as.
The transmission pattern is the input of the model, and the output is either the house dwellers are present at home or absent.

% For simplicity, it is expected that there are frequent power consumption changes when the dwellers are at home, which results in more transmitted packets due to the CAT approach. Analyzing the transmission pattern and observing that the number of packets is significantly lower than that when the dwellers are at home, it is highly possible that they are not at home.

%%%%%%%%%%%%%%%%%%%%%%%%%%%%%%%%%%%%%%%%%%%%%%%%%%%%%%%%
\subsubsection{Training the Attacker's Model}\label{Att_training}
% Although the attacker can use any machine learning model to launch PPA, 
% A CNN-based model is used by the attacker to capture the temporal correlation in the input sequence; however, the attacker can use other types of machine-learning models.
A CNN-based model is used by the attacker to capture the temporal correlation in the input sequence.
While training the attacker's model, $\ell2-$regularization is used to limit over-fitting. An overfitted model is a model that performs well on the training data and it does not perform well on new/unseen data. Also, we adjust the hyper-parameters of the CNN-based attacker model using hyperopt tool~\cite{Bergstra_2015} on a validation dataset to tune the number of units in each layer, batch size, learning rate, and select an activation function for each layer.
Moreover, \textit{Adam}~\cite{kingma2019method} optimizer is used to train the model for 60 epochs, 128 batch size, 0.001 learning rate, and categorical cross entropy as the loss function. To train the attacker's model, we used Python3 libraries such as Scikit-learn~\cite{scikit-learn}, Numpy, TensorFlow~\cite{tensorflow2015-whitepaper}, and Keras~\cite{chollet2015keras}, which are installed on a high-performance cluster (HPC) of the Tennessee Tech University using a NVIDIA Tesla K80 GPU. 
The best performance of the attacker's model can be achieved using the hyper-parameters given in Table~\ref{tab:CNN_arch}.

% Note that, despite the fact that recurrent neural networks (RNNs) can also be used to detect temporal correlations in the input sequence, but it gives a very close performance compared with the CNN model. However, RNNs suffers from slow convergence~\cite{kani2017drrnn} which means that many iterations are needed to reach the final solution and corresponds to an increase in the computational time.
% However, RNNs CNNs tend to be much slower (\~5 times slower) than CNN.

\subsubsection{Attacker's Deep-Learning Model Architecture}
In the following, we present the detailed architecture of the attacker's CNN-based model.
% to infer whether a consumer is present or absent by oberving his/her transmission pattern. 
1-dimension CNN model (Conv1D) is used which consists of four convolution layers, an average pooling layer, four fully connected layers, and a softmax output layer.
Table~\ref{tab:CNN_arch} gives the detailed structure of the attacker's model in case of using 1/5min and 1/30min transmission rates including the number of layers, hidden units, and activation functions. The structure can be described as follows.

{\renewcommand{\arraystretch}{1.35}
\begin{table}[t]
\centering
\caption{Attacker's model architecture, where AF stands for activation function, and $h_i$ is the $i^{th}$ fully connected hidden layer.}
\label{tab:CNN_arch}
\resizebox{1\columnwidth}{!}{%
\begin{tabular}{| >{\centering\arraybackslash}m{1.7cm}|
>{\centering\arraybackslash}m{1.27cm}|
>{\centering\arraybackslash}m{1.27cm}| >{\centering\arraybackslash}m{1.27cm}|
>{\centering\arraybackslash}m{1.27cm}|}
\hline
\rowcolor[gray]{0.8}
\cellcolor[gray]{0.8}                      & \multicolumn{2}{c|}{\cellcolor[gray]{0.8}No. of units using} & \multicolumn{2}{c|}{\cellcolor[gray]{0.8}AF using} \\ 
\hhline{|>{\arrayrulecolor[gray]{0.8}}-|>{\arrayrulecolor{black}}----|}
\rowcolor[gray]{0.8} 
\multirow{-2}{*}{\cellcolor[gray]{0.8}Layer} &  1/5min  & 1/30min & 1/5min  & 1/30min  \\ \hline
\hhline{|>{\arrayrulecolor{black}}-|>{\arrayrulecolor{black}}-|>{\arrayrulecolor{black}}-|>{\arrayrulecolor{black}}-|}
Input      &  288   & 48 &  Linear & Linear\\ \hline
Conv1D     &  150   &80 & ELu & ReLU\\ \hline
Conv1D      &  85  & 32&  ReLU  & ReLU\\ \hline
Conv1D      &  45 &  20&  ReLU  & ELu\\ \hline
Conv1D      &  25 &  --&  ReLU & --\\ \hline
MaxPooling1D       &  2 &2  &  -- & --\\ \hline
$h_1$      &  512 & 256 & ELu & ELu\\ \hline
$h_2$      &  512  &512&  ReLU &  Sigmoid\\ \hline
$h_3$      &  128 & 64&  Sigmoid & ELu\\ \hline
$h_4$      &  64 &  64& ELu & ReLU\\ \hline
Output     & 2   &   2&  Softmax & Sigmoid\\ \hline
\end{tabular}}
\end{table}}

% {\renewcommand{\arraystretch}{1.35}
% \begin{table}[t]
% \centering
% \caption{Attacker's model architecture, where AF stands for activation function, and $h_i$ is the $i^{th}$ fully connected hidden layer.}
% \label{tab:CNN_arch}
% \begin{tabular}{| >{\centering\arraybackslash}m{2.2cm}| >{\centering\arraybackslash}m{2.2cm}| >{\centering\arraybackslash}m{2.2cm}|}
% \hline \rowcolor[gray]{0.8}
% \textbf{Layer} & \cellcolor[gray]{0.8}  \textbf{No. of units} & \cellcolor[gray]{0.8}  \textbf{AF} \\ \hline
%  Input      &  288    &  Linear\\ \hline
%  Conv1D     &  150    &  Elu \\ \hline
%  Conv1D      &  85   &  ReLU  \\ \hline
%  Conv1D      &  45   &  Elu  \\ \hline
%  Conv1D      &  25   &  Sigmoid \\ \hline
%  MaxPooling1D       &  2   &  -- \\ \hline
%  $h_1$      &  512   &  Elu \\ \hline
%  $h_2$      &  128   &  ReLU \\ \hline
%  $h_3$      &  64  &  Sigmoid \\ \hline
%  $h_4$      &  64   &  Elu \\ \hline
%  Output     & 2      &  Softmax\\ \hline
% \end{tabular}
% \end{table}}

%--------------------------------------------------
\begin{figure*}[t]
\centering
\includegraphics[width=6.8in]{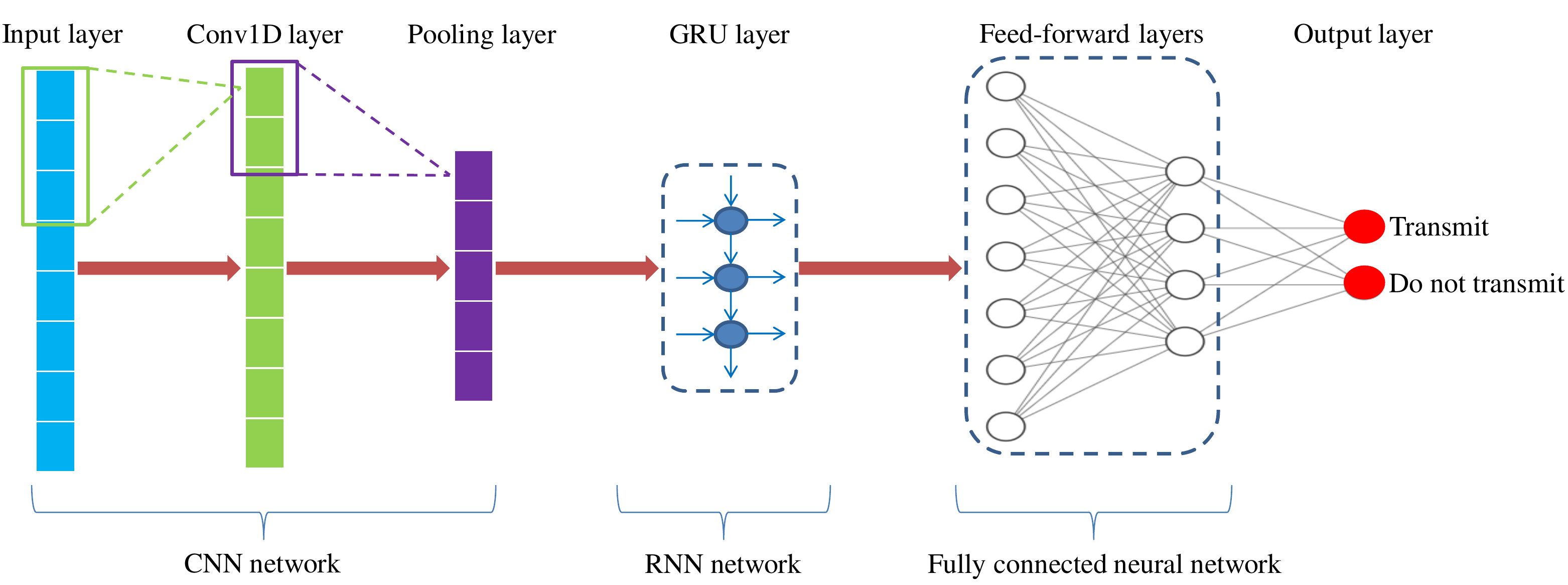}
\caption{Illustration of the CNN-RNN-based defense architecture.} \label{RSPDL_Model}
\end{figure*}

%-------------------------------------------------------

\begin{itemize}
    \item The input layer is the first layer of the model, and it consists of nodes, called neurons. It receives the input data and passes them to the following layers. The number of neurons in the input layer is equal to the number of readings per day (or transmission pattern size), e.g., 288 readings per day if the transmission rate is one reading every 5 minutes.

    \item 1-dimension convolution layers are used to capture the temporal correlation in the input sequence using a number of convolution filters. 
    
    \item A max pooling layer is used to combine the features extracted from the previous layer. 
    
    \item Fully connected hidden layers are used to extract more features and make complex decisions.
    
    \item Softmax output layer is used to classify whether the dwellers are present or absent. Since we only have two classes, the output layer has two units. 
\end{itemize}

%$\hat{X}_{\text{\sc{tst}}}$. 
%%%%%%%%%%%%%%%%%%%%%%%%%%%%%%%%%%%%%%%%%%%%%%%%%%%%%%%%
\subsection{Defense Model}

The objective of this paper is to thwart transmission analysis attacks, including PPA, caused by using the CAT approach to preserve the consumers' privacy while using their fine-grained power consumption readings for load monitoring and energy management. No one, including eavesdroppers, the utility, and the aggregator, should be able to learn whether the house dwellers are present at home or absent. The key challenge is how to send spoofing transmissions so that the attacker cannot distinguish between the transmission pattern generated by our defense model when a consumer is absent and the transmission pattern when the consumer is at home. The attacker should not also identify the spoofing transmissions because they are sent only when the consumer is absent.
Therefore, in this subsection, we explain how the defense model is trained to generate spoofing transmissions using deep-learning to countermeasure PPA. 
% A time series refers to a sequence of equally spaced data points that represent the evolution of a specific variable, i.e., decision, over time. The model decision is enabled through modeling the dependencies between the current decision and transmission observations.

\subsubsection{Dataset preparation for training the defense model}
We used the generated dataset in Section~\ref{dataset_preprocessed} to create another dataset to train the defense model as follows. 
% extracted from the records when the house dwellers are present at home.
First, we create a sliding window of size $n$ elements and slide it on the transmission pattern of the consumer, so the new dataset contains $n$ elements from the transmission pattern and the label is the next element in the pattern. Note that each element in the transmission pattern is either one or zero in case there is a transmission or no transmission, respectively, in the corresponding time slot.
% transform the daily transmission patterns when the consumers are present at home into sliding windows with a predetermined number of look-back transmission observations, which are one if there is a transmission event and zero if there is no transmission event, and a look-ahead of size $1$. Thus, the features/input of the resulted dataset contains $n$ transmission observations and its label is the $(n+1)^{th}$ observation. 
In other words, the input of the defense model is a sequence of $n$ binary features, zeros and ones, and the label is the transmission decision in the next time slot, where one corresponds to transmit and zero corresponds to no transmit.
By this way, our defense model acts as a predictor that takes $n$ elements from a transmission pattern and predicts next element. Since the model is trained on transmission patterns when the consumer is present at home, the model outputs an element that makes the pattern looks like the patterns generated when the consumer is present at home.

%--------------------------------------------------

\subsubsection{Defense model architecture}
In the following, we present the detailed architecture of the defense model. As shown in Fig.~\ref{RSPDL_Model}, the defense model is a hybrid deep-learning model that combines a CNN with GRU followed by a fully connected neural network and a Softmax output layer. 
The input of the defense model is the transmission pattern of the last $n$ time slots and the output is the decision of either the SM needs to send a redundant power consumption reading (spoofing transmission) or not so that the generated transmission pattern looks like the patterns transmitted when the house dwellers are present at home. The same configuration, mentioned in Section~\ref{Att_training}, that is used in training the attacker's model, is also used in training the defense model, but with a batch size of 400 and the learning rate of 0.0001. The best performance of the defense model can be achieved using the hyper-parameters given in Table~\ref{tab:RSPDL_arch}.

The main reason of using a deep-learning model is that comparing with the traditional shallow learning methods such as logic regression, support vector machine, decision tree, etc., it can better capture the complex patterns within the data, which can improve the performance of the model. Moreover, the combination of CNN and GRU improves the extraction of the correlations between the previous transmission pattern and the current transmission decision, which is the key to improve the decision accuracy.
% ~\cite{7804912}

% shallow learning is often ineffective in dealing with complex problems, even when the number of samples is sufficient. This is because the network complexity is low and the learning ability is not strong. its network depth is deeper, its nonlinear representation ability is stronger, and its network learning ability is stronger. The combination of GRU and CNN can better capture the temporal and spatial correlation in the input sequence.

{\renewcommand{\arraystretch}{1.35}
\begin{table}[t]
\centering
\caption{Defense model structure, where AF stands for activation function, and $h_i$ is the $i^{th}$ fully connected hidden layer.}
\label{tab:RSPDL_arch}
\resizebox{1\columnwidth}{!}{%
\begin{tabular}{| >{\centering\arraybackslash}m{1.7cm}|
>{\centering\arraybackslash}m{1.27cm}|
>{\centering\arraybackslash}m{1.27cm}| >{\centering\arraybackslash}m{1.27cm}|
>{\centering\arraybackslash}m{1.27cm}|}
\hline
\rowcolor[gray]{0.8}
\cellcolor[gray]{0.8}                      & \multicolumn{2}{c|}{\cellcolor[gray]{0.8}No. of units using} & \multicolumn{2}{c|}{\cellcolor[gray]{0.8}AF using} \\ 
\hhline{|>{\arrayrulecolor[gray]{0.8}}-|>{\arrayrulecolor{black}}----|}
\rowcolor[gray]{0.8} 
\multirow{-2}{*}{\cellcolor[gray]{0.8}Layer} &  1/5min  & 1/30min & 1/5min  & 1/30min  \\ \hline
\hhline{|>{\arrayrulecolor{black}}-|>{\arrayrulecolor{black}}-|>{\arrayrulecolor{black}}-|>{\arrayrulecolor{black}}-|}
Input      &  100   & 35 &  Linear & Linear\\ \hline
Conv1D     &  150   &128 & ReLU & ReLU\\ \hline
Conv1D     &  --   &64 & -- & ReLU\\ \hline
Conv1D     &  --   &32 & -- & ReLU\\ \hline
MaxPooling1D       &  4 & 2  &  -- & --\\ \hline
GRU       &  200 & 128  & Tanh & Tanh\\ \hline
$h_1$      &  128 & 128 & ReLU & ReLU\\ \hline
$h_2$      &  32  & 32 &  ReLU &  ReLU\\ \hline
Output     & 2   &   2&  Softmax & Softmax\\ \hline
\end{tabular}}
\end{table}}

%--------------------------------------------------------------------
\begin{figure}[t]
\centering
\includegraphics[width=3.5in]{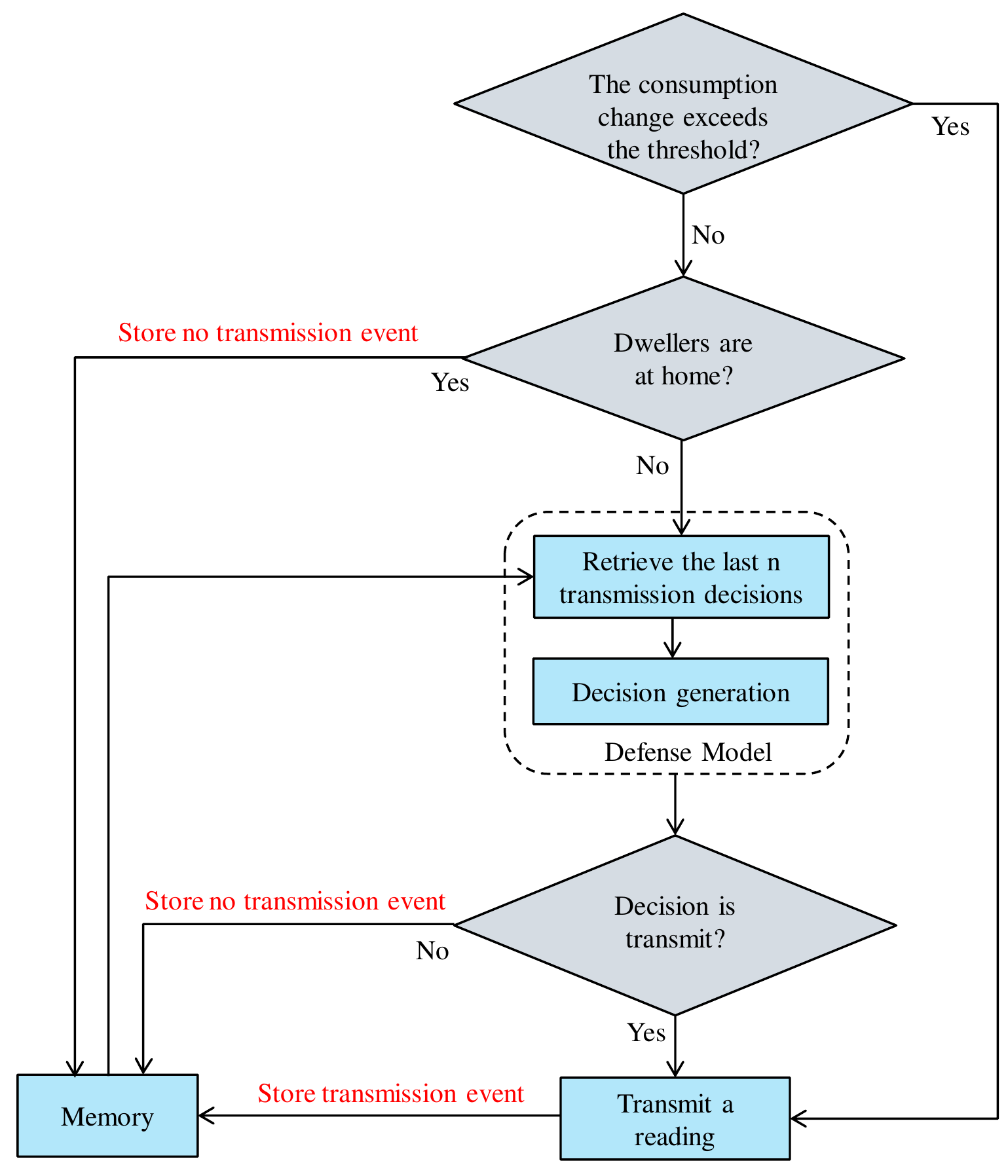}
\caption{An illustration of the defense methodology.} \label{RSPDL}
\vspace{-0.1in}
\end{figure}
%--------------------------------------------------------------------
\subsubsection{Defense methodology}\label{Defense_methodology}
Fig.~\ref{RSPDL} shows the defense methodology that should be run in each time slot when the consumer is absent to decide whether a spoofing transmission (redundant reading) should be sent to defend against PPA.
When the absolute value of the percentage of the change in the power consumption exceeds a threshold, regardless of the presence of dwellers, the SM should send the updated power consumption reading to the aggregator. On the other hand, when the dwellers are not at home, they run the defense model to decide whether the SM should transmit redundant readings to prevent the attacker from distinguishing the generated pattern by the defense model from the transmission patterns transmitted when the consumer is present at home to preserve the consumers' privacy. A memory, as shown in Fig.~\ref{RSPDL}, is used to store the transmission decision of each time slot, and it needs to keep only the last $n$ transmission decisions. 
% Therefore, a sliding window is used to discard the old transmission decisions. 

Our defense model works as follows.
First, it retrieves the last $n$ transmission decisions from the SM's memory and inputs them to the defense model, which is presented in Fig.~\ref{RSPDL_Model}. 
Then, the model outputs a decision which is either ``transmit'' or ``do not transmit'' a redundant power consumption reading (a spoofing transmission) as shown in Fig.~\ref{RSPDL_Model}. 
We have tried different values of $n$ and we found that the defense model gives good results at $n=100$ and $n=35$ in case of using 1/5min and 1/30min transmission rates, respectively.

\subsection{Privacy-preserving Reading Collection Using CAT Approach}
\subsubsection{System Initialization}
An offline KDC should bootstrap the system as follows. First, it generates the Paillier cryptosystem's public key $(n=$ $p q, g),$ and the corresponding private key $(\lambda, \mu)$ where $p$ and $q$ are two large prime numbers with $|p|=|q| .$ Moreover, it generates the bilinear pairing parameters $\left(q_{1}, \mathbb{G}, \mathbb{G}_{T}, P, \hat{e}\right)$. Furthermore, it chooses a secure hash function $H,$ where $H$: $\{0,1\}^{*} \rightarrow \mathbb{G}.$ Finally, it publishes the system parameters as $p u b s=\left\{n, g, q_{1}, \mathbb{G}, \mathbb{G}_{T}, P, \hat{e}, H\right\}$. In addition, each $SM_i$ chooses a secret key $x_{i} \in \mathbb{Z}_{q}^{*}$, and computes the corresponding public key $Y_{i}=x_{i} P$. Similarly, the aggregator possesses private/public key pairs $x_{g w} / Y_{g w}$, respectively.

\subsubsection{Reporting the Power Consumption Readings by SMs}
When an $SM_i$ transmits a reading either due to a change in the consumption or a redundant reading (spoofing transmission) to preserve privacy, it encrypts the power consumption reading $m_{i}$ and sends it to the aggregator by performing the following steps.

\begin{itemize}

    \item Step 1: $SM_i$ chooses a random number $r_{i} \in \mathbb{Z}_{n}^{*}$ and encrypts $m_{i}$ using Paillier cryptosystem as follows.
$$
C_{i}=g^{m_{i}} \cdot r_{i}^{n} \quad \bmod n^{2}
$$

\item Step 2: $SM_i$ uses its private key $x_{i}$ to generate a signature $\sigma_{i}$ as
$$
\sigma_{i}=x_{i} H\left(C_{i} \| TS\right),
$$
where $T S$ is a timestamp
\item Step 3: $SM_i$ sends the following tuple to the aggregator.
$$
C_{i}\left\|TS\right\|\sigma_{i}
$$
\end{itemize}
\subsubsection{Aggregation by the Aggregator}
The aggregator should store the last encrypted power consumption reading sent by each SM to reuse it in case that the SM does not send an encrypted reading because there is no enough change in the consumption. 
% In other words, when sending the spoofing transmissions, they should be a real time encrypted readings not a null transmissions to be able to aggregated at the aggregator. 
% due to the CAT approach or RSPDL. On the other hand, it should update the stored reports list by replacing the stored report with the new one. Also, if there is no stored report for $SM_i$, the aggregator should add a new entry for $SM_i$ to the list.
The aggregator should do the following steps to verify the SMs' messages and calculate the ciphertext of the total aggregated reading of all SMs.

\begin{itemize}
    \item Step $1$: Verification of the received messages: After receiving messages from the SMs, the aggregator first checks whether the timestamps are fresh or not. Then, it uses a batch verification approach to efficiently verify the received signatures instead of verifying the signatures individually as follows, where $w$ is the number of messages and $w \leq |\mathbb{SM}|$.
$$
\hat{e}\left(\sum_{i=1}^{w} \sigma_{i}, P\right) \stackrel{?}{=} \prod_{i=1}^{w} \hat{e}\left(H\left(C_{i} \| T S\right), Y_{i}\right)
$$
\textbf{Proof of signature verification: } \\
	\begin{align*}
	\hat{e}\left(\sum_{i=1}^{w} \sigma_{i}, P\right) &\stackrel{}{=} \prod_{i=1}^{w} \hat{e}\left(\sigma_{i}, P\right)\\
	&\stackrel{}{=} \prod_{i=1}^{w} \hat{e}\left(x_{i} H\left(C_{i} \| TS\right), P\right)\\
	&\stackrel{}{=}\prod_{i=1}^{w} \hat{e}\left(H\left(C_{i} \| TS\right), x_{i} P\right)\\
	&\stackrel{}{=}\prod_{i=1}^{w} \hat{e}\left(H\left(C_{i} \| TS\right), Y_{i}\right).
   \end{align*}
   
% The aggregator moves to the next step if the batch verification process passes, otherwise, it should verify the individual signatures and drop the report that has an invalid signature.

    \item Step $2$: Privacy-preserving readings aggregation: The aggregator should compute the encrypted aggregated fine-grained reading $C_{g w}$ as follows.
        $$
        C_{g w}=\prod_{i=1}^{|\mathbb{SM}|} C_{i} \quad \bmod n^{2}
        $$
    
    \item Step $3$: The aggregator uses its private key $x_{g w}$ to compute the signature as follows.
        $$
        \sigma_{g w}=x_{g w} H\left(C_{g w} \| T S\right)
        $$
    \item Step $4$: The aggregator composes a message containing the encrypted aggregated reading and a signature and sends it to the EU. The message should have the following tuple.
        $$
        C_{g w}\|T S\| \sigma_{g w}
        $$

\end{itemize}

\subsubsection{Aggregated Reading Recovery by the EU}
Upon receiving the message from the aggregator, the EU checks the freshness of the timestamp and verifies the signature by checking the following.
$$
\hat{e}\left(\sigma_{g w}, P\right) \stackrel{?}{=} \hat{e}\left(H\left(C_{g w} \| T S\right), Y_{g w}\right)
$$

\textbf{Proof of signature verification: }
	\begin{align*}
	\hat{e}(\sigma_{g w}, P) &\stackrel{}{=} \hat{e}(x_{g w} H(C_{g w} \| TS), P)\\
	&\stackrel{}{=} \hat{e}(H(C_{g w} \| TS), x_{g w} P)\\
	&\stackrel{}{=}\hat{e}\left(H\left(C_{g w} \| T S\right), Y_{g w}\right).
   \end{align*}
    
Then, the EU uses the secret key $(\lambda, \mu)$ to decrypt $C_{g w}$, and recover the total power consumption reading of the SMs $R_{T}$ as follows.
$$
R_{T}=D\left(C_{g w}\right)=L\left(C_{g w}^{\lambda} \bmod n^{2}\right) \cdot \mu \bmod n 
  =\sum_{i=1}^{|\mathbb{SM}|} m_{i}
$$

\section{Evaluations}
\label{PE}

In this section, we first discuss the performance metrics and then evaluate the attacker's success rate in determining the absence of the house dwellers by analyzing the transmission pattern without using our defense. Next, we evaluate the success of the attacker while using our defense model assuming that the attacker does not know the defense model. Finally, we evaluate the attacker's success assuming that the attacker knows the defense model.

% if the attacker does not have knowledge about how the spoofing transmissions are generated. Then, we test the robustness of STDL in case that the attacker is aware of the defense methodology.

\subsection{Performance Metrics} In order to evaluate our models' performance, we consider the following metrics. The success rate ($SR$) is the probability that the attacker successfully detects that the house dwellers are not at home (absent).
The false alarm rate ($FA$) is the probability that the attacker thinks that the dwellers are absent but they are actually present at home. 
% The third metric is the accuracy that measures the percentage of absent/present consumers' records that are correctly detected as absent/present. 
The defense model performance is better when $SR$ is low. These metrics are measured as follows.

\begin{equation*}
    \small
   \textrm{$SR$} =  \frac{\textrm{$TP$}}{\textrm{$TP$} + \textrm{$FP$}}, \quad \textrm{$FA$} = \frac{\textrm{$FP$}}{\textrm{$TN$} + \textrm{$FN$}},
\end{equation*}
% \begin{equation*}
%     \small
%  \quad  Accuracy = \frac{\textrm{$TP$}+ \textrm{$TN$}}{\textrm{$TN$} + \textrm{$TP$} + \textrm{$FP$} + \textrm{$FN$}}, 
%   \end{equation*}
where, $TP$, $TN$, $FN$, and $FP$ stand for true positive, true negative, false negative, and false positive, respectively. Moreover, we measure also the efficiency of the CAT approach with and without our defense in terms of the percentage of readings that are not transmitted, which is related to the saving in the communication bandwidth, compared to continuous transmission approach. The efficiency is computed as follows.
\begin{equation*}
    \small
   \textrm{Efficiency (\%)} =  \frac{\textrm{$P_t$}-\textrm{$R_t$}}{\textrm{$P_t$}}\times100,
   \end{equation*}
where $P_t$ is the number of readings that are sent when SMs send the readings continuously (without using CAT approach) and $R_t$ is the number of readings sent by our scheme that uses the CAT approach.

\subsection{Attacker's Success Rate Without Our Scheme}\label{without_ourscheme}

As mentioned in Section~\ref{subsec:Threat Model}, the attacker has  old absent/present records of the transmission patterns of the consumers. These records are used to train a deep-learning model, which is discussed in Section~\ref{att_model}, to infer the absence of house dwellers by running the model using the consumer's transmission pattern. Using this model and without using our defense model, the attacker's success rates are 91.1\% and 88.5\% in case of 1/5min and 1/30min transmission rates, respectively, as can be seen in Table~\ref{Performance_evaluation}. The reduction in the success rate with the increasing of the transmission interval is because as the transmission interval increases, the number of transmissions decreases, which results in hiding detailed information from the consumption pattern. 
% reduces the model's ability to capture the correlation of the input, and this results in lower success. 
From the given results, we can conclude that the attacker can infer the absence of the dwellers with high confidence.

\subsection{Attacker's Success Rate With Our Scheme if the Defense Model is Unknown}\label{not_known}
In this subsection, we evaluate the performance of our defense model assuming that the attacker does not know the defense model. The attacker uses the transmission patterns generated by our defense model as input to the attacker model, discussed in Section~\ref{att_model}, to know whether the house dwellers are absent.
% Since the STDL defense scheme works only when the house dwellers are not at home, our defense scheme is evaluated when the dwellers are absent. 
As shown in Table~\ref{Performance_evaluation}, the attacker's success rate is significantly reduced from 91.1\%, without using our defense model, to only 3.15\% using our defense with 1/5min transmission rate, while the attacker's success rate is reduced from 88.5\% to 2.25\% with 1/30min transmission rate.
This big reduction in the success rate of the attacker indicates that our defense model can generate patterns that look like the patterns transmitted when the consumer is at home, and that is why the attacker's model was not able to classify the patterns generated by our defense model as absent.
% This is because STDL is based on a hybrid deep learning approach that exploits the potential of using CNN+RNN neural networks in better capturing the complex correlations between the previous transmission observations and the current decision.

Moreover, using 1/5min transmission rate at $ 10\%$ threshold, our scheme can achieve efficiency of 40.8\% compared to 69.67\% in case that privacy is not preserved. Also, using 1/30min transmission rate, the efficiency is reduced from 44.6\% to 19.01\% using our defense, as shown in Table~\ref{Performance_evaluation}.
This reduction in efficiency is due to sending redundant readings to preserve privacy, and this can be considered the cost for privacy preservation. 
Note that better efficiency can be achieved by reducing the number of redundant readings but with higher attacker's success, i.e., there is a trade off between the efficiency and attacker's success rate.
% because more spoofing transmissions (which reduce the efficiency) are sent to achieve less success rate. 
From our experiment results, we can see that the less the transmission rate (longer transmission interval), the more reduction in efficiency since the likelihood that the power consumption change exceeds the threshold increases which causes more transmissions.

{\renewcommand{\arraystretch}{1.35}
\begin{table}[t]
\centering
\caption{Evaluations with and without our defense model for 1/5min and 1/30min transmission rates at $ 10\%$ threshold.}
\label{Performance_evaluation}
\resizebox{1\columnwidth}{!}{%
\begin{tabular}{| >{\centering\arraybackslash}m{3.6cm}| >{\centering\arraybackslash}m{1.8cm}| >{\centering\arraybackslash}m{1.8cm}|}
\hline \rowcolor[gray]{0.8}
\textbf{Transmission rate} & \cellcolor[gray]{0.8}  \textbf{1/5min} & \cellcolor[gray]{0.8}  \textbf{1/30min} \\ \hline \hline
 Attacker's success rate without our scheme      &  91.1\%    &  88.5\%\\ \hline
 Attacker's success rate with our scheme    &  \textcolor{blue}{\textbf{3.15\%}}    &  \textcolor{blue}{\textbf{2.25\%}} \\ \hline \hline
 Efficiency without our scheme      &  69.67\%   &  44.6\%  \\ \hline
 Efficiency with our scheme     &  \textcolor{blue}{\textbf{40.8\%}}   &  \textcolor{blue}{\textbf{19.01\%}}  \\ \hline
\end{tabular}}
\end{table}}

%%%%%%%%%%%%%%%%%%%%%%%%%%%%%%%%%%%%%%%%%%%%%%%%%%%%%%%%%%%%%%%%%

\subsection{Attacker's Success With Our Scheme if the Defense Model is Known}

In this subsection, we evaluate the performance of our scheme when the attacker knows the defense model. 
In Sections~\ref{without_ourscheme} and~\ref{not_known}, we assume that the attacker has old absent/present records of the transmission patterns of the consumers, and using these records, the attacker trained a deep-learning model to learn whether a consumer is absent. In this subsection, we assume a stronger attacker who knows not only old transmission patterns but also the defense model.
Using the available knowledge to the attacker, he can launch powerful attacks as follows. First, the attacker uses the absent records and the defense model to create ``spoofing'' patterns. Then, using absent, present, and spoofing patterns, the attacker trains an advanced model to classify the input patterns into ``absent'', ``present'', or ``spoofing''. Hence, if the classification is ``spoofing'', the attacker thinks that the victim consumer is absent and the defense model is used to generate patterns with spoofing transmissions.

The best performance of the attacker's model, when the defense model is known in case of using 1/5min and 1/30min transmission rates, can be achieved using the hyper-parameters given in Table~\ref{att_CNN_arch}.
By knowing the defense model by the attacker, the success rate of detecting the absence of the house dwellers increases from 3.15\% to 13.52\% and from 2.25\% to 12.84\% using 1/5min and 1/30min transmission rates, respectively. 
These results are good given the strong assumptions that the attacker knows the defense model and has old transmission patterns.
% Then, the attacker trains another deep learning model to be able to differentiate the spoofing transmissions which indicate that the absence of the house dwellers.
% Since the attacker has  absent/present records in addition to having the defense model, he/she can obtain the spoofing generated transmission patterns of the absent records.
% if some one says that why we do not compare with literature in this table, the answer will be it is difficult to compare since we use different data, different solution (our solution is based on ML while Li's is based on heuristic), and Li's did not compute some metrics such as savings.

{\renewcommand{\arraystretch}{1.35}
\begin{table}[t]
\centering
\caption{Attacker's model architecture when the defense model is known to the attacker, where AF stands for activation function, and $h_i$ is the $i^{th}$ fully connected hidden layer.}
\label{att_CNN_arch}
\resizebox{1\columnwidth}{!}{%
\begin{tabular}{| >{\centering\arraybackslash}m{1.7cm}|
>{\centering\arraybackslash}m{1.27cm}|
>{\centering\arraybackslash}m{1.27cm}| >{\centering\arraybackslash}m{1.27cm}|
>{\centering\arraybackslash}m{1.27cm}|}
\hline
\rowcolor[gray]{0.8}
\cellcolor[gray]{0.8}                      & \multicolumn{2}{c|}{\cellcolor[gray]{0.8}No. of units using} & \multicolumn{2}{c|}{\cellcolor[gray]{0.8}AF using} \\ 
\hhline{|>{\arrayrulecolor[gray]{0.8}}-|>{\arrayrulecolor{black}}----|}
\rowcolor[gray]{0.8} 
\multirow{-2}{*}{\cellcolor[gray]{0.8}Layer} &  1/5min            & 1/30min            & 1/5min            & 1/30min  \\ \hline
\hhline{|>{\arrayrulecolor{black}}-|>{\arrayrulecolor{black}}-|>{\arrayrulecolor{black}}-|>{\arrayrulecolor{black}}-|}
Input      &  288   & 48 &  Linear & Linear\\ \hline
Conv1D     &  150   &128 & ReLU & ReLU\\ \hline
MaxPooling1D       &  4 &4  &  -- & --\\ \hline
GRU     &  --   &32 & -- & Tanh\\ \hline
$h_1$      &  128 & 64 & ReLU & ReLU\\ \hline
$h_2$      &  35  &32&  ReLU &  ReLU\\ \hline
Output     & 3   &   3&  Softmax & Softmax\\ \hline
\end{tabular}}
\end{table}}

% Tables~\ref{confusion_matrix_kown_5} and~\ref{confusion_matrix_kown_30} give the confusion matrix when the attacker knows the defense model using 1/5min and 1/30min transmission rates, respectively, at $ 10\%$ threshold.
% The given results in Table~\ref{confusion_matrix_kown_5} indicate that the attacker's success in identifying the spoofing patterns generated by our defense model is only 13.52\%.
% % , while most of the absent records are detected as present, 86.48\%. 
% While the results in Table~\ref{confusion_matrix_kown_30} support the ones presented in Table~\ref{confusion_matrix_kown_5}, and using 1/30min transmission rate, the attacker's success in identifying the spoofing patterns generated by our defense model is only 12.84\%.
% % , and 85.36\% of the absent records are detected as present. 
% This proves the efficiency of our defense methodology using different transmission rates.

\begin{figure}[t]
\centering
\includegraphics[width=3.4in]{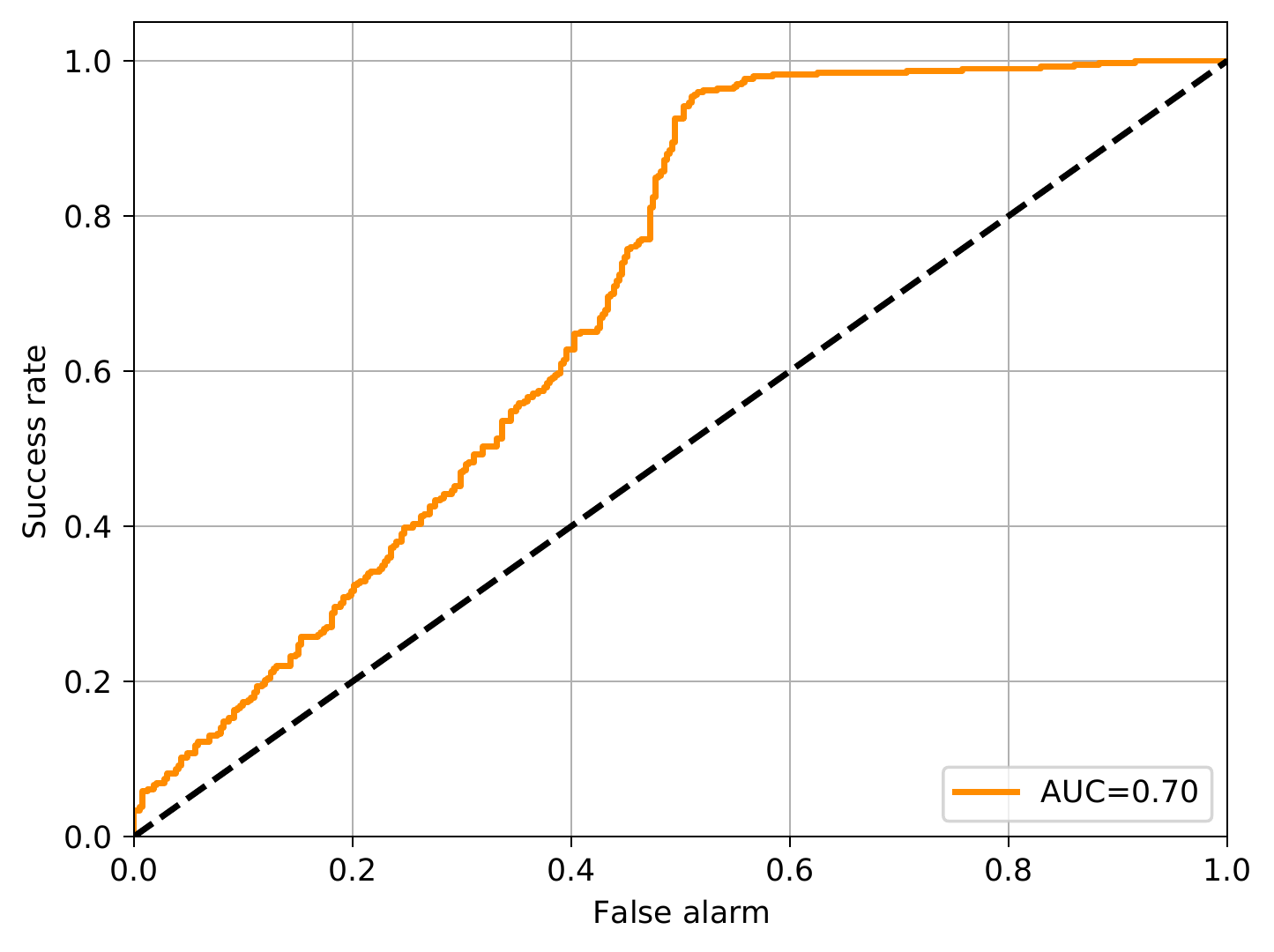}
\caption{The ROC curve of the attacker model when the defense model is known using 1/5min transmission rate at $ 10\%$ threshold.} \label{roc_5_10_known}
\end{figure}

\begin{figure}[t]
\centering
\includegraphics[width=3.4in]{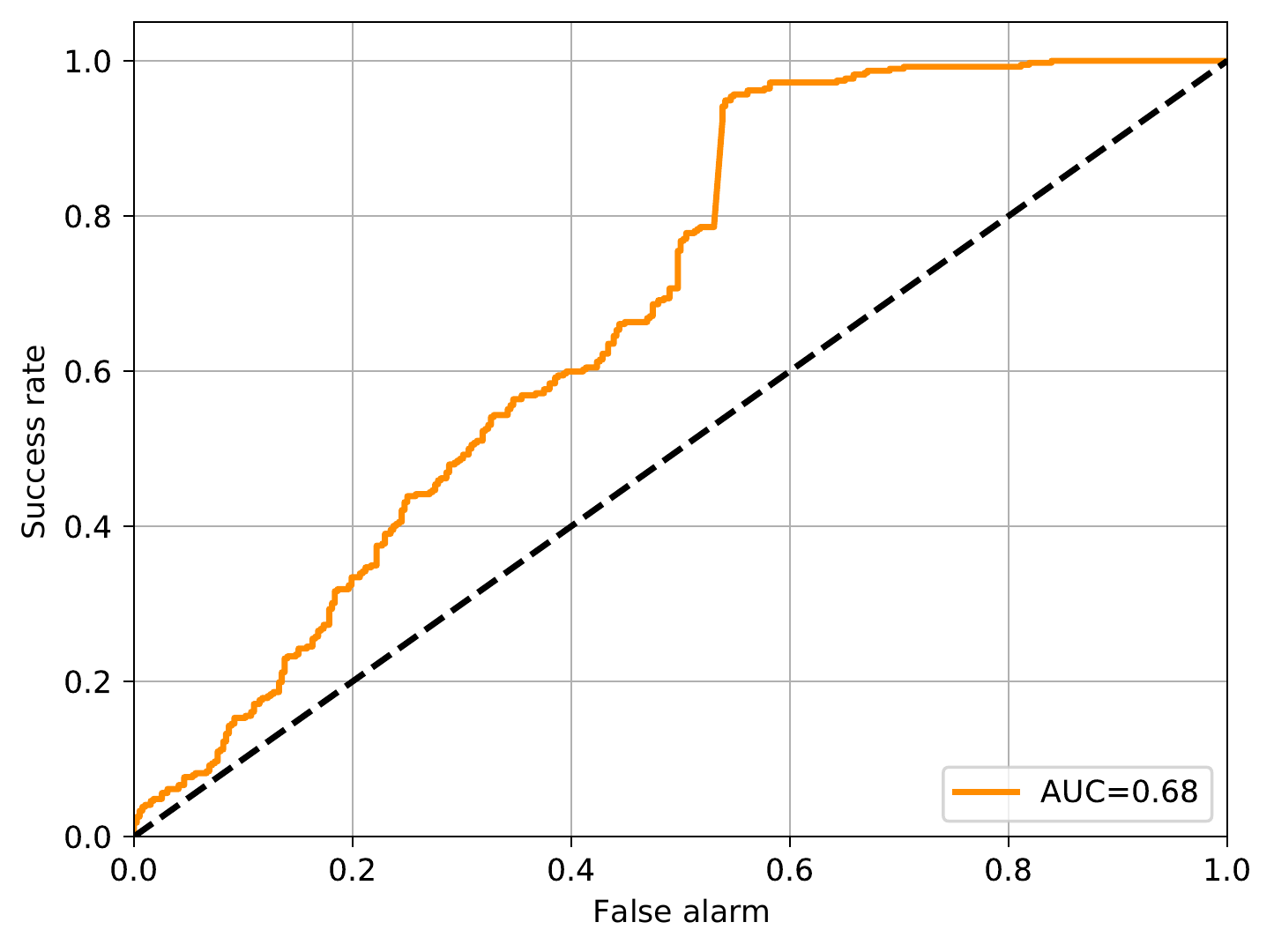}
\caption{The ROC curve of the attacker model when the defense model is known using 1/30min transmission rate at $ 10\%$ threshold.} \label{roc_30_10_known}
\end{figure}

In addition, Figs.~\ref{roc_5_10_known} and~\ref{roc_30_10_known} show the Receiver Operating Characteristic (ROC) curves of the attacker model in case of 1/5min and 1/30min transmission rates, respectively, assuming that the defense model is known. 
ROC curve is often used to evaluate the classification accuracy, which is measured by the area under the curve (AUC). This area indicates how the model can distinguish between the classes, where a bigger AUC indicates a better performance.
% Note that the main concern of the attacker is the false alarm since a false alarm may cause significant damage to the attacker.
We assume that a false alarm rate of 0.05 is acceptable to the attacker, and then, we need to determine the success rate at this false alarm rate.
As can be seen in Figs.~\ref{roc_5_10_known} and~\ref{roc_30_10_known}, at false alarm of 0.05, the attacker's success rate is fairly low (less than 0.1) compared to more than 0.6 using ASP proposed in~\cite{Dr_Li}. 
Therefore, the attacker can detect the absence of the house dwellers with a high probability in ASP~\cite{Dr_Li}, while the success rate in our scheme is much lower.
This is because our scheme is trained on the transmission patterns when the consumer is present at home, and then it uses spoofing transmissions (redundant real readings) when the consumer is absent to generate a pattern that looks similar to the consumer's transmission patterns when he/she is at home.

\subsection{Discussions}\label{Security_Analysis}
In this paper, we consider a strong attack model assuming that the attacker has old transmission patterns and knows the defense model. Using this knowledge, the attacker trains a sophisticated deep-learning model to detect the absence of the house dwellers. 
On the other hand, the design of our defense scheme is based on a deep-learning model which is trained on the transmission patterns when the consumers are present at home to generate spoofing transmissions so that the attacker cannot distinguish the pattern generated using STDL from the patterns transmitted when the consumers are at home to preserve consumers' privacy.
% Moreover, the design of our defense scheme is based on a deep learning approach, which usually performs better than heuristic-based approaches~\cite{jokar2016electricity,9148937}.
% Moreover, unlike~\cite{Dr_Li}, our defense model is general, and can be applied to new consumers. 
The scheme in~\cite{Dr_Li} is customized, i.e., the scheme depends on the distribution of the power consumption changes (transmission patterns) of each consumer, but our model is general (one model for all consumers) and it can be applied to new consumers who have no old transmission patterns.
Furthermore, we assume that the network nodes, including the SMs, EU, and aggregator, are honest but curious, while the scheme in~\cite{Dr_Li} assumes that they are trusted and the defense is limited only to the eavesdroppers.

% In addition, compared to the proposed defense scheme in~\cite{Dr_Li}, our defense methodology only requires a small memory to store the previous few transmission observations of the consumer(e.g., the last 100 observations). While ASP~\cite{Dr_Li} needs a high memory to store a large number of history templates to be used in generating the spoofing transmissions.

Moreover, our scheme can achieve the following desirable privacy requirements that can counter the attacks mentioned in Section~\ref{subsec:Threat Model}. 

\textbf{PPA countermeasure:} An adversary can eavesdrop on the wireless transmissions between the SMs and the aggregator and construct the transmission patterns of the consumers, and then launch PPA. 
On the other hand, the aggregator and EU are curious to learn whether the house dwellers are absent by analyzing their SMs' transmission patterns.
However, using our scheme, no one can learn whether the house dwellers are absent because it sends spoofing transmissions (redundant real readings) to make it difficult for the attacker to learn this information.

% STDL uses previous observations of the consumers' transmissions to predict future decisions, i.e., whether the SM needs to send a redundant power consumption reading or not, so that the generated transmission pattern looks like as the house dwellers are present at home. 

% \textcolor{red}{The aggregator should not be able to distinguish spoofing transmissions from the real ones, and it does not need to be able to differentiate between them, since the aggregator should use the received encrypted readings for aggregation even if the readings are sent as redundant readings (spoofing transmissions).}

\textbf{Consumers' fine-grained readings' privacy preservation:} To ensure the privacy of the consumers, each SM first encrypts its readings using the public key of the Paillier cryptosystem and no entity is able to learn the individual readings to preserve consumers' privacy. In addition, if the same reading is repeated at different times, the ciphertext looks different because the encryption uses a one-time random number. 
Also, when an SM sends a redundant reading that is similar to the last reported reading to preserve privacy, the ciphertext looks different and the aggregator cannot know that the reading is redundant. This is important to prevent the aggregator and eavesdroppers from identifying redundant readings (spoofing transmissions) because these readings are sent only when the house dwellers are absent.
It is important to use the random number $r$ once because if it is reused, the difference between two readings, i.e., $m_1$ and $m_2$, can be obtained by dividing their ciphertexts $g^{m_1} \cdot r^{n} /g^{m_2} \cdot r^{n} = g^{m_1-m_2}$, and this can be solved to obtain ${m_1-m_2}$; hence, by knowing one reading, the other one can be obtained. 
% Therefore, it is essential to not reuse the random number. To learn a certain consumer's power consumption reading, the EU must collude with ($\mathbb{|SM|}$-1) consumers. This can be done by subtracting the total power consumption of the colluding SMs from the total power consumption known to the EU. This attack is not feasible when the number of SMs in an AMI network is large.

\textbf{Confidentiality of AMI's total power consumption:} The aggragator should send the ciphertext of the AMI's total power consumption to the EU for load monitoring, after computing it from the encrypted fine-grained power consumption readings of the SMs.
The aggregator and the attackers, who eavesdrop on the communications between the aggregator and the EU learn nothing about the total consumption of an AMI since the private key which is known only to the EU is needed for decrypting Paillier ciphertext to obtain the aggregated power consumption reading.

\section{Conclusion}\label{conclusion}
In this paper, we have proposed STDL, an efficient and privacy-preserving scheme for collecting power consumption readings in AMI networks. Our scheme is efficient because the SMs do not send the readings if there is no enough change in the power consumption comparing to the last report reading. It also preserves the consumers' privacy by transmitting redundant readings (spoofing transmissions) using a deep-learning-based defense scheme.
% We also used homomorphic encryption scheme to hide the consumers' power consumption readings while allowing the EU to obtain the aggregated reading for load monitoring.
First, a real dataset for power consumption readings dataset and clustering technique are used to create a dataset for transmission patterns using CAT approach.
Then, we train an attacker model using deep-learning, and our evaluations indicate that the attacker's success rate is about 91\%. Next, we train a deep-learning-based defense scheme to send spoofing transmissions efficiently to thwart PPA attacks.
Extensive evaluations are conducted, and the results indicate that our scheme can reduce the attacker's success rate to 13.52\% in case he knows the defense model and to 3.15\% in case he does not know the model, while still achieving high efficiency in terms of the number of readings that should be transmitted.
This big reduction in the attacker's success rate indicates that our defense can generate patterns that look like the patterns transmitted when the consumer is at home.
Moreover, our measurements indicate that the proposed scheme can reduce the number of readings that should be transmitted by about 41\% while preserving the privacy of the consumers.

% Unlike~\cite{Dr_Li}, our defense model is general (i.e., one model for all consumers) that does not rely on specific consumer's behaviour/distribution, and can be applied to new consumers who have no history of power consumption.

\section*{Acknowledgement}
This project was funded by the Deanship of Scientific Research (DSR) at King Abdulaziz University, Jeddah, under grant no. DF-745-611-1441. The authors, therefore, acknowledge with thanks DSR for technical and financial support.

\IEEEpeerreviewmaketitle

\newcolumntype{P}[1]{>{\centering\arraybackslash}p{#1}}
%\balance
\bibliographystyle{IEEEtran}
\bibliography{main}

% Generated by IEEEtran.bst, version: 1.14 (2015/08/26)
\begin{thebibliography}{10}
\providecommand{\url}[1]{#1}
\csname url@samestyle\endcsname
\providecommand{\newblock}{\relax}
\providecommand{\bibinfo}[2]{#2}
\providecommand{\BIBentrySTDinterwordspacing}{\spaceskip=0pt\relax}
\providecommand{\BIBentryALTinterwordstretchfactor}{4}
\providecommand{\BIBentryALTinterwordspacing}{\spaceskip=\fontdimen2\font plus
\BIBentryALTinterwordstretchfactor\fontdimen3\font minus
  \fontdimen4\font\relax}
\providecommand{\BIBforeignlanguage}[2]{{%
\expandafter\ifx\csname l@#1\endcsname\relax
\typeout{** WARNING: IEEEtran.bst: No hyphenation pattern has been}%
\typeout{** loaded for the language `#1'. Using the pattern for}%
\typeout{** the default language instead.}%
\else
\language=\csname l@#1\endcsname
\fi
#2}}
\providecommand{\BIBdecl}{\relax}
\BIBdecl

\bibitem{6298960}
V.~C. {Gungor}, D.~{Sahin}, T.~{Kocak}, S.~{Ergut}, C.~{Buccella}, C.~{Cecati},
  and G.~P. {Hancke}, ``A survey on smart grid potential applications and
  communication requirements,'' \emph{IEEE Transactions on Industrial
  Informatics}, vol.~9, no.~1, pp. 28--42, Feb. 2013.

\bibitem{5741147}
Z.~M. {Fadlullah}, M.~M. {Fouda}, N.~{Kato}, A.~{Takeuchi}, N.~{Iwasaki}, and
  Y.~{Nozaki}, ``Toward intelligent machine-to-machine communications in smart
  grid,'' \emph{IEEE Communications Magazine}, vol.~49, no.~4, pp. 60--65, Apr.
  2011.

\bibitem{5622050}
C.~Efthymiou and G.~Kalogridis, ``Smart grid privacy via anonymization of smart
  metering data,'' \emph{Proc. of IEEE international conference on smart grid
  communications}, pp. 238--243, Oct. 2010.

\bibitem{issue1}
G.~W. Hart, ``Nonintrusive appliance load monitoring,'' \emph{Proceedings of
  the IEEE}, vol.~80, no.~12, pp. 1870--1891, 1992.

\bibitem{7841782}
H.~{Mohammed}, S.~{Tonyali}, K.~{Rabieh}, M.~{Mahmoud}, and K.~{Akkaya},
  ``Efficient privacy-preserving data collection scheme for smart grid {AMI}
  networks,'' \emph{Proc. of IEEE Global Communications Conference (GLOBECOM)},
  pp. 1--6, Dec. 2016.

\bibitem{ISNCC20}
M.~I. Ibrahem, M.~M. Badr, M.~M. Fouda, M.~Mahmoud, W.~Alasmary, and Z.~M.
  Fadlullah, ``{PMBFE}: Efficient and privacy-preserving monitoring and billing
  using functional encryption for {AMI} networks,'' \emph{Proc. of IEEE
  International Symposium on Networks, Computers and Communications (ISNCC'20),
  Montreal, Canada}, Oct. 2020.

\bibitem{ibrahem2020efficient}
M.~I. {Ibrahem}, M.~{Nabil}, M.~M. {Fouda}, M.~{Mahmoud}, W.~{Alasmary}, and
  F.~{Alsolami}, ``Efficient privacy-preserving electricity theft detection
  with dynamic billing and load monitoring for {AMI} networks,'' \emph{IEEE
  Internet of Things Journal}, 2020, doi: 10.1109/JIOT.2020.3026692.

\bibitem{Dr_Li}
H.~{Li}, S.~{Gong}, L.~{Lai}, Z.~{Han}, R.~C. {Qiu}, and D.~{Yang}, ``Efficient
  and secure wireless communications for advanced metering infrastructure in
  smart grids,'' \emph{IEEE Transactions on Smart Grid}, vol.~3, no.~3, pp.
  1540--1551, Sep. 2012.

\bibitem{5403146}
M.~A. {Lisovich}, D.~K. {Mulligan}, and S.~B. {Wicker}, ``Inferring personal
  information from demand-response systems,'' \emph{IEEE Security Privacy},
  vol.~8, no.~1, pp. 11--20, Feb. 2010.

\bibitem{7093120}
S.~{Finster} and I.~{Baumgart}, ``Privacy-aware smart metering: A survey,''
  \emph{IEEE Communications Surveys Tutorials}, vol.~17, no.~2, pp. 1088--1101,
  Apr. 2015.

\bibitem{weaver2014perspective}
\BIBentryALTinterwordspacing
K.~Weaver, ``{A perspective on how smart meters invade individual privacy},''
  Last accessed: Nov. 2020. [Online]. Available:
  \url{https://skyvisionsolutions.files.wordpress.com/2014/08/utility-smart-meters-invade-privacy-22-aug-2014.pdf}
\BIBentrySTDinterwordspacing

\bibitem{7066595}
S.~{Yussof}, M.~E. {Rusli}, Y.~{Yusoff}, R.~{Ismail}, and A.~A. {Ghapar},
  ``Financial impacts of smart meter security and privacy breach,'' \emph{Proc.
  of International Conference on Information Technology and Multimedia}, pp.
  11--14, Nov. 2014.

\bibitem{6177682}
M.~{Alam}, M.~{Reaz}, and M.~{Ali}, ``A review of smart homes—past, present,
  and future,'' \emph{IEEE Transactions on Systems, Man, and Cybernetics, Part
  C (Applications and Reviews)}, vol.~42, no.~6, pp. 1190--1203, Nov. 2012.

\bibitem{6521385}
K.~{Samarakoon}, J.~{Ekanayake}, and N.~{Jenkins}, ``Reporting available demand
  response,'' \emph{IEEE Transactions on Smart Grid}, vol.~4, no.~4, pp.
  1842--1851, Dec. 2013.

\bibitem{CARRIEARMEL2013213}
K.~{Carrie Armel}, A.~Gupta, G.~Shrimali, and A.~Albert, ``Is disaggregation
  the holy grail of energy efficiency? the case of electricity,'' \emph{Energy
  Policy}, vol.~52, pp. 213 -- 234, Jan. 2013.

\bibitem{7362875}
S.~{Werner} and J.~{Lunden}, ``Event-triggered real-time metering in smart
  grids,'' \emph{Proc. of European Signal Processing Conference (EUSIPCO)}, pp.
  2701--2705, Sep. 2015.

\bibitem{7229338}
S.~{Werner} and J.~{Lundén}, ``Smart load tracking and reporting for real-time
  metering in electric power grids,'' \emph{IEEE Transactions on Smart Grid},
  vol.~7, no.~3, pp. 1723--1731, May. 2016.

\bibitem{7007667}
J.~{Lunden} and S.~{Werner}, ``Real-time smart metering with reduced
  communication and bounded error,'' \emph{Proc. of IEEE International
  Conference on Smart Grid Communications (SmartGridComm)}, pp. 326--331, Nov.
  2014.

\bibitem{7875140}
M.~{Simonov}, G.~{Chicco}, and G.~{Zanetto}, ``Event-driven energy metering:
  Principles and applications,'' \emph{IEEE Transactions on Industry
  Applications}, vol.~53, no.~4, pp. 3217--3227, Aug. 2017.

\bibitem{7479541}
A.~{Proano}, L.~{Lazos}, and M.~{Krunz}, ``Traffic decorrelation techniques for
  countering a global eavesdropper in {WSNs},'' \emph{IEEE Transactions on
  Mobile Computing}, vol.~16, no.~3, pp. 857--871, Mar. 2017.

\bibitem{7289106}
S.~{Alsemairi} and M.~{Younis}, ``Adaptive packet-combining to counter traffic
  analysis in wireless sensor networks,'' \emph{Proc. of International Wireless
  Communications and Mobile Computing Conference (IWCMC)}, pp. 337--342, Aug.
  2015.

\bibitem{pairing}
D.~Boneh and M.~Franklin, ``Identity-based encryption from the weil pairing,''
  \emph{Advances in Cryptology --- CRYPTO 2001}, pp. 213 -- 229, 2001.

\bibitem{paillier1999public}
P.~Paillier, ``Public-key cryptosystems based on composite degree residuosity
  classes,'' \emph{Proc. of international conference on the theory and
  applications of cryptographic techniques}, pp. 223--238, Apr. 1999.

\bibitem{zheng2018wide}
Z.~{Zheng}, Y.~{Yang}, X.~{Niu}, H.~{Dai}, and Y.~{Zhou}, ``Wide and deep
  convolutional neural networks for electricity-theft detection to secure smart
  grids,'' \emph{IEEE Transactions on Industrial Informatics}, vol.~14, no.~4,
  pp. 1606--1615, Apr. 2018.

\bibitem{10.5555/1213811}
S.~Haykin, \emph{Neural Networks and Learning Machines: A Comprehensive
  Foundation (3rd Edition)}.\hskip 1em plus 0.5em minus 0.4em\relax USA:
  Prentice-Hall, Inc., Nov. 2008.

\bibitem{lecun1995convolutional}
Y.~LeCun \emph{et~al.}, ``Convolutional networks for images, speech, and time
  series,'' \emph{The handbook of brain theory and neural networks}, vol. 3361,
  no.~10, pp. 255--258, 1995.

\bibitem{ha2018recurrent}
D.~Ha and J.~Schmidhuber, ``Recurrent world models facilitate policy
  evolution,'' \emph{in Advances in Neural Information Processing Systems}, pp.
  2450--2462, 2018.

\bibitem{clustering}
K.~M. Faraoun and A.~Boukelif, ``Neural networks learning improvement using the
  k-means clustering algorithm to detect network intrusions,''
  \emph{International Journal of Computer and Information Engineering}, vol.~1,
  no.~10, pp. 3151 -- 3158, 2007.

\bibitem{6745416}
S.~{Chaturvedi}, R.~N. {Titre}, and N.~{Sondhiya}, ``Review of handwritten
  pattern recognition of digits and special characters using feed forward
  neural network and {Izhikevich} neural model,'' \emph{Proc. of International
  Conference on Electronic Systems, Signal Processing and Computing
  Technologies}, pp. 425--428, Jan. 2014.

\bibitem{kingma2019method}
D.~P. Kingma and J.~Ba, ``{Adam}: {A} method for stochastic optimization,''
  \emph{Proc. of International Conference on Learning Representations, San
  Diego, CA, USA}, May. 2015.

\bibitem{Bergstra_2015}
J.~Bergstra, B.~Komer, C.~Eliasmith, D.~Yamins, and D.~D. Cox, ``Hyperopt: a
  {Python} library for model selection and hyperparameter optimization,''
  \emph{Computational Science {\&} Discovery, doi:
  https://doi.org/10.1088/1749-4699/8/1/014008}, 2015.

\bibitem{8545748}
M.~{Nabil}, M.~{Ismail}, M.~{Mahmoud}, M.~{Shahin}, K.~{Qaraqe}, and
  E.~{Serpedin}, ``Deep recurrent electricity theft detection in {AMI} networks
  with random tuning of hyper-parameters,'' \emph{Proc. of International
  Conference on Pattern Recognition (ICPR)}, pp. 740--745, Aug 2018.

\bibitem{yin2017comparative}
W.~Yin, K.~Kann, M.~Yu, and H.~Sch{\"u}tze, ``Comparative study of {CNN} and
  {RNN} for natural language processing,'' \emph{arXiv preprint
  arXiv:1702.01923}, 2017.

\bibitem{8985885}
A.~F. {Ganai} and F.~{Khursheed}, ``Predicting next word using {RNN} and {LSTM}
  cells: Stastical language modeling,'' \emph{Proc. of International Conference
  on Image Information Processing (ICIIP)}, pp. 469--474, Nov. 2019.

\bibitem{6638947}
A.~{Graves}, A.~{Mohamed}, and G.~{Hinton}, ``Speech recognition with deep
  recurrent neural networks,'' \emph{Proc. of IEEE International Conference on
  Acoustics, Speech and Signal Processing}, pp. 6645--6649, May. 2013.

\bibitem{dash2010hybridized}
B.~Dash, D.~Mishra, A.~Rath, and M.~Acharya, ``A hybridized k-means clustering
  approach for high dimensional dataset,'' \emph{International Journal of
  Engineering, Science and Technology}, vol.~2, no.~2, pp. 59--66, 2010.

\bibitem{nwankpa1811activation}
C.~Nwankpa, W.~Ijomah, A.~Gachagan, and S.~Marshall, ``Activation functions:
  Comparison of trends in practice and research for deep learning,''
  \emph{arXiv preprint arXiv:1811.03378}, 2018.

\bibitem{PCRdataset}
\BIBentryALTinterwordspacing
Kolter, ``{Residential Energy Disaggregation Dataset (REDD)},'' Last accessed:
  2020. [Online]. Available:
  \url{http://traces.cs.umass.edu/index.php/Smart/Smart}
\BIBentrySTDinterwordspacing

\bibitem{scikit-learn}
F.~Pedregosa \emph{et~al.}, ``Scikit-learn: Machine learning in {P}ython,''
  \emph{Journal of Machine Learning Research}, vol.~12, pp. 2825--2830, Oct.
  2011.

\bibitem{tensorflow2015-whitepaper}
\BIBentryALTinterwordspacing
A.~Mart\'{\i}n \emph{et~al.}, ``{TensorFlow}: Large-scale machine learning on
  heterogeneous systems,'' 2015, software available from tensorflow.org.
  [Online]. Available: \url{https://www.tensorflow.org/}
\BIBentrySTDinterwordspacing

\bibitem{chollet2015keras}
F.~Chollet \emph{et~al.}, ``Keras,'' \url{https://github.com/fchollet/keras},
  2015.

\end{thebibliography}

\end{document}